\documentclass[prb,preprint]{revtex4-1}

\usepackage{graphics}
\usepackage{amsmath}
\usepackage{amsfonts}
\usepackage{graphicx} % needed for figures
\usepackage{amsmath,esint} % use math package
\usepackage{multirow}
\usepackage{color}

\def\rcurs{{\mbox{$\resizebox{.11in}{.05in}{\includegraphics{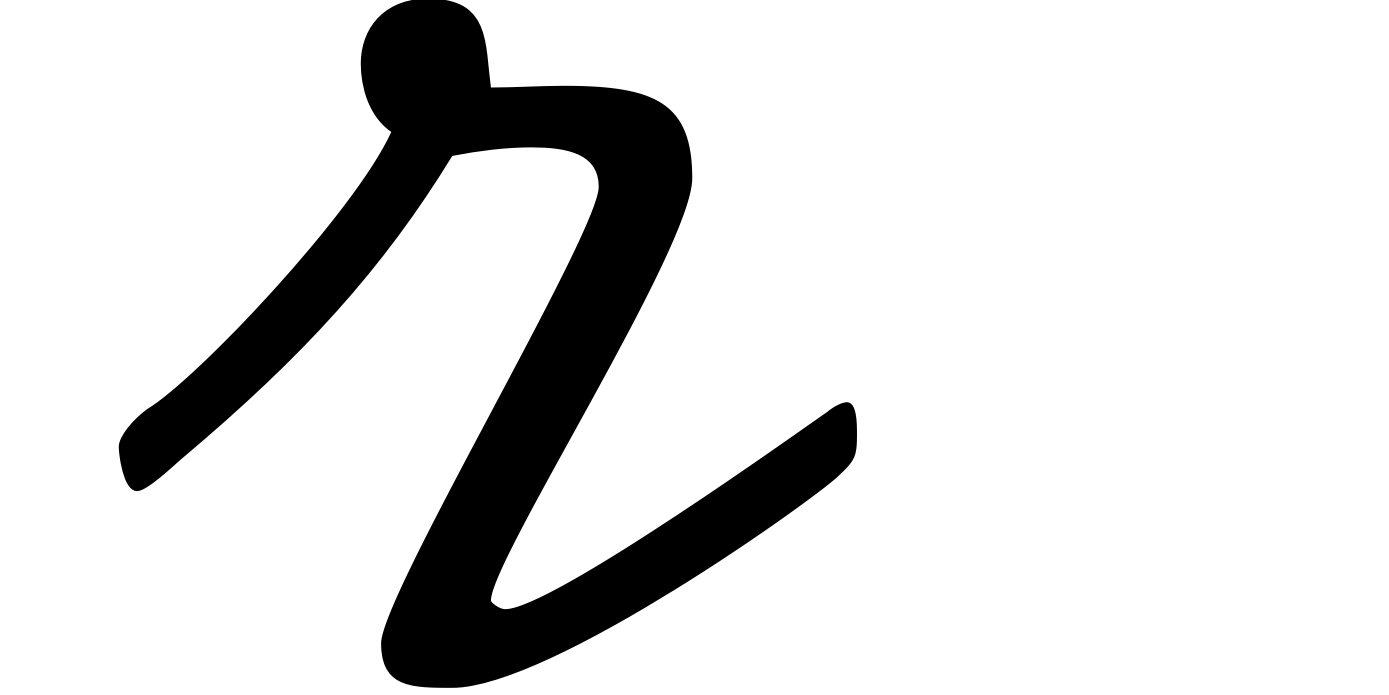}}$}}}
\def\lrcurs{{\mbox{$\resizebox{.14in}{.075in}{\includegraphics{ScriptR}}$}}}
\def\brcurs{{\mbox{$\resizebox{.11in}{.05in}{\includegraphics{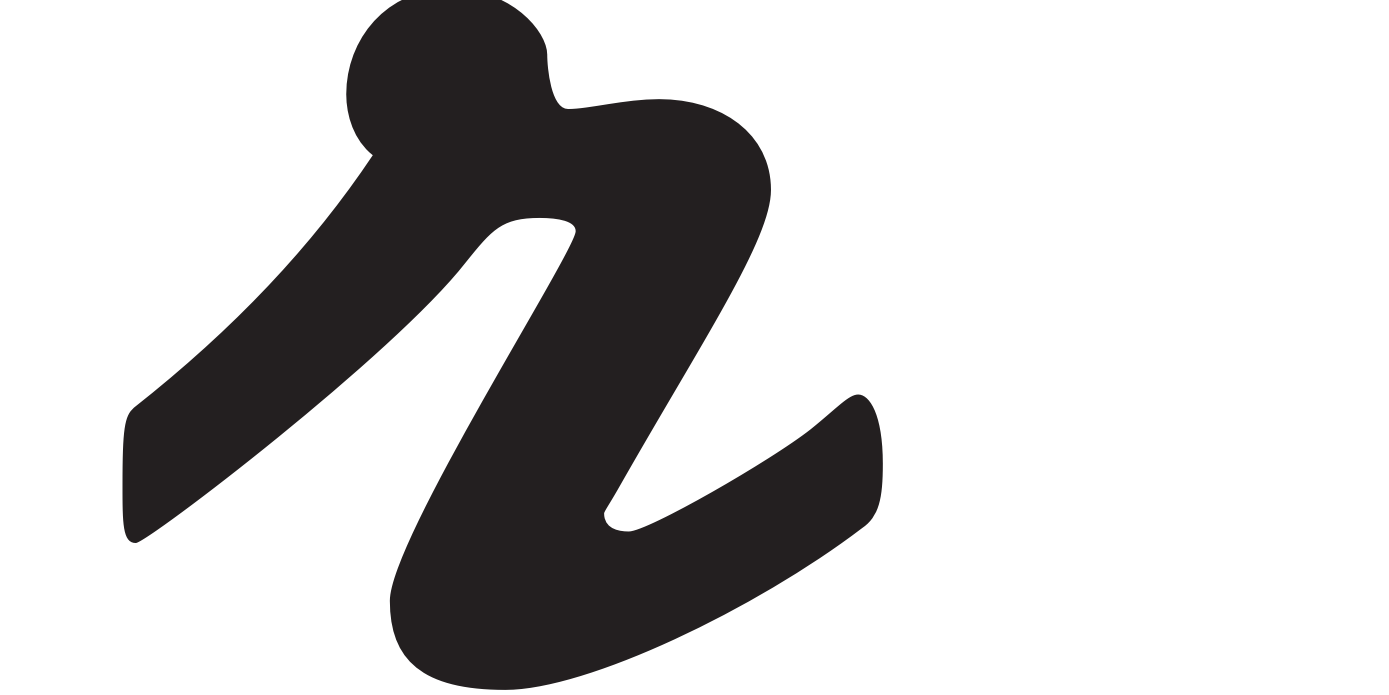}}$}}}
\def\lbrcurs{{\mbox{$\resizebox{.14in}{.075in}{\includegraphics{BoldR}}$}}}

\def\lhrcurs{{\mbox{$\hat \lbrcurs$}}}

\begin{document}

\title{Divergence of electric field of continuous and of a point charge for relativistic and non-relativistic motion}
\author{Altay Zhakatayev}
\email{azhakatayev@nu.edu.kz} % optional

\affiliation{azhakatayev@nu.edu.kz, School of Science and Technology, Nazarbayev University, Astana 010000, Kazakhstan}

\date{\today}

\begin{abstract}
In this paper we considered divergence of electric and of magnetic fields for four cases: classical point charge, classical continuous charge, relativistic point and relativistic continuous charges. Results for classical and relativistic point charges are the same as in literature, i.e. Gauss's law is valid. However results for time-varying classical and relativistic distributed charges indicate that divergence of electric field is not zero even for volumes of space where no charges are present. For these cases original Gauss's law might require modification. Divergence of electric field seems to be far-field type scalar anisotropic field, which is generated by time-varying electric charges or currents. Results indicate that for these effects to be sufficiently large to be experimentally observable the time variation of electric charges and/or of currents should be very fast. Divergence of magnetic field is zero for all cases. 
\end{abstract}

\maketitle

\section{Introduction}\label{sec:introduction}

Maxwell equations describe relations between electric and magnetic fields and electric charges
\begin{subequations}\label{eqn:Maxwell}
	\begin{align}
	\nabla\cdot\mathbf{E}&=\frac{\rho}{\varepsilon_{0}}\label{eqn:Gauss}\\
	\nabla\cdot\mathbf{B}&=0 \label{eqn:Gauss2}\\
	\nabla\times{\mathbf{E}}&=-\frac{\partial\mathbf{B}}{\partial{t}} \label{eqn:Faraday}\\
	\nabla\times{\mathbf{B}}&=\mu_{0}(\varepsilon_{0}\frac{\partial{\mathbf{E}}} {\partial{t}} + \mathbf{J}) \label{eqn:Maxwell2}
	\end{align}
\end{subequations}
where $\mathbf{E}$ and $\mathbf{B}$ are electric and magnetic force field strengths, $\mathbf{J}$ and $\rho$ are electric current and electric charge densities, $\varepsilon_{0}$ and $\mu_{0}$ are permittivity and permeability of space, $\cdot$ and $\times$ denote dot and cross product.  Using (\ref{eqn:Gauss2}) and vector calculus identity that divergence of curl of arbitrary continuous vector field is zero, magnetic vector potential field $\mathbf{A}$ is introduced:
\begin{equation}\label{eqn:vector_potential}
\mathbf{B} = \nabla \times \mathbf{A}
\end{equation}
If (\ref{eqn:vector_potential}) is substituted into (\ref{eqn:Faraday}) and vector calculus identity, that curl of gradient of arbitrary continuous scalar field is zero, is used, then electric scalar potential field $V$ is introduced:
\begin{equation}\label{eqn:scalar_potential}
\mathbf{E} = -\nabla V - \frac{\partial{\mathbf{A}}}{\partial{t}}
\end{equation}
Even though $\mathbf{E}$ and $\mathbf{B}$ are force fields, while $V$ and $\mathbf{A}$ are potential fields, for simplicity we will refer to the former as \emph{fields}, while to the latter as \emph{potentials}. Historically electric and magnetic fields were discovered (or introduced), and therefore were used, before the vector and scalar potentials. However there is a point of view, supported even by founders of classical electrodynamics, like Maxwell and Faraday, that these vector and scalar potentials are more fundamental than electric and magnetic fields \cite{DavidGriffiths}. This assumption becomes apparent when we derive Li\'enard-Wiechert potentials and fields for a point charge. For example, in his classical textbook ``Introduction to Electrodynamics" Professor D. Griffiths argues that for a point and continuous charges it is correct first to extend the classical potentials to relativistic motion and then from these extended potentials to find relativistic fields. If we try simply to extend the non-relativistic fields to relativistic case, then we get the wrong result. In other words, classical electrodynamic fields lead to classical electrodynamic potentials, which are extended to relativistic electrodynamic potentials, which in turn lead to relativistic fields. This relationship between electrodynamic potentials and fields is often encountered in science. If a fact or event $X$ in all cases leads to $Y$ ($X\rightarrow Y$), it doesn't necessarily mean that $Y$ in all cases leads to $X$ ($Y\not\rightarrow X$). In other words, if $Y$ is more fundamental than $X$, i.e. it is generalization or extension of $X$, then $Y$ only in special cases leads to $X$. In this work we simply want to go one step further and check if divergence of electric and of magnetic fields, calculated from electrodynamic potentials, still satisfies the Gauss's law. For example, if it turns out that electrodynamic fields, obtained from potentials, do not satisfy the Gauss's law, then it would mean that potentials are more fundamental than fields. If, however, fields do satisfy the Gauss's law, then it would mean that potentials are not more fundamental than fields. 

Using (\ref{eqn:vector_potential}) and (\ref{eqn:scalar_potential}) four Maxwell equations in (\ref{eqn:Maxwell}) can be reduced to the following two equations \cite{DavidGriffiths}
\begin{subequations}\label{eqn:Maxwell3}
	\begin{align}
	\square^2 V + \frac{\partial{L}}{\partial{t}} &= -\frac{\rho}{\varepsilon_{0}}\\
	\square^2 \mathbf{A} - \nabla L &= -\mu_0 \mathbf{J}	
	\end{align}
\end{subequations}
where $L$ is defined as
\begin{equation}\label{eqn:Lorentz_gauge}
L = \nabla \cdot \mathbf{A} + \mu_0 \varepsilon_{0} \frac{\partial{V}}{\partial{t}}
\end{equation}
while $\square = \nabla^2 - \mu_0 \varepsilon_{0} \frac{\partial^2}{\partial{t^2}}$ denotes d'Alembertian operator. For a vector valued function $\mathbf{A}$ Laplacian operator can be written as $\nabla^2 \mathbf{A} = \nabla (\nabla \cdot \mathbf{A}) - \nabla \times (\nabla \times \mathbf{A})$.
Often using gauge transformation, $L$ is set to $0$, which is called Lorentz (Lorenz) gauge condition. For simplicity we will refer to $L$ as \emph{Lorentz gauge function}. In Lorentz gauge condition (\ref{eqn:Maxwell3}) become 
\begin{subequations}\label{eqn:Maxwell_Lorentz}
	\begin{align}
	\square^2 V &= -\frac{\rho}{\varepsilon_{0}}\\
	\square^2 \mathbf{A} &= -\mu_0 \mathbf{J}	
	\end{align}
\end{subequations}
In this work we also will check if Lorentz gauge condition is satisfied for electrodynamic potentials for different cases.

\section{Theoretical Methodology}\label{sec:theoretical_methodology}

In order to calculate the divergence of fields, we have to first calculate the fields themselves. However as mentioned above, fields are obtained from scalar and vector potentials \cite{DavidGriffiths}. Scalar and vector potentials are different for point and for distributed charges. For a point and/or distributed charge, potentials also differ depending on relativistic or non-relativistic cases. Non-relativistic case is often called \emph{classical}. In relativistic motion we can not ignore retardation effects, or we have to consider finite speed of information propagation in the field. While in classical case retardation effects can be ignored, or in other words the speed of propagation of information can be taken as infinite. Therefore in this paper we will consider four cases: classical and relativistic point charge, classical and relativistic continuous (distributed) charge. Table \ref{tab:potentials} summarizes and presents the four different cases of electrodynamic potentials \cite{DavidGriffiths,WolfgangPanofsky}. From mathematical point of view, these vector potentials represent solution of non-homogeneous wave equations in (\ref{eqn:Maxwell_Lorentz}) \cite{JerroldFranklin,WalterGreiner}. This will be our starting point. For these four different cases, we will re-derive electric and magnetic fields, divergences of these fields and we will check whether Lorenz gauge condition holds for them, i.e. if Lorentz gauge function is zero ($L=0$) and if (\ref{eqn:Maxwell_Lorentz}) are satisfied. 

\begin{table}
\begin{tabular}{ c  c | c | c |}
	\cline{3-4}
	& & \multicolumn{2}{c|}{Speed} \\
	\cline{3-4}
	& & Classical & Relativistic \\
	& & ($v \ll c$) & ($v \sim c$) \\	
	\hline
	\multicolumn{1}{ |c }{\multirow{6}{*}{Charge}} &
	\multicolumn{1}{ |c| }{Point} & & \\
	\multicolumn{1}{ |c| }{} & & $V(\mathbf{r},t) = \frac{q}{4\pi \varepsilon_0 \rcurs}$ & $V(\mathbf{r},t) = \frac{qc}{4\pi \varepsilon_0 (\rcurs c - \brcurs \cdot \mathbf{v})} $\\
	\multicolumn{1}{ |c| }{} & & $\mathbf{A}(\mathbf{r},t) = \frac{\mu_0 \mathbf{I}}{4\pi \rcurs}$ & $\mathbf{A}(\mathbf{r},t) = \frac{\mu_0 qc \mathbf{v}}{4\pi (\rcurs c - \brcurs \cdot \mathbf{v})}$ \\
	\multicolumn{1}{ |c| }{} & & & \\	
	\cline{2-4}
	\multicolumn{1}{ |c }{} &
	\multicolumn{1}{ |c| }{Distributed} & & \\
	\multicolumn{1}{ |c| }{} & & $V(\mathbf{r},t) = \frac{1}{4\pi \varepsilon_0} \int \frac{\rho(\mathbf{r}',t_r)}{\rcurs} d\tau'$ & $V(\mathbf{r},t) = \frac{c}{4\pi \varepsilon_0} \int \frac{\rho(\mathbf{r}',t_r)}{(\rcurs c - \brcurs \cdot \mathbf{v})} d\tau'$\\
	\multicolumn{1}{ |c| }{} & & $\mathbf{A}(\mathbf{r},t) = \frac{\mu_0}{4\pi} \int \frac{\mathbf{J}(\mathbf{r}',t_r)}{\rcurs} d\tau'$ & $\mathbf{A}(\mathbf{r},t) = \frac{\mu_0c}{4\pi} \int \frac{\mathbf{J}(\mathbf{r}',t_r)}{(\rcurs c - \brcurs \cdot \mathbf{v})} d\tau'$ \\
	\multicolumn{1}{ |c| }{} & & & \\
	\hline                                                      
	 
\end{tabular}
\caption{Electrodynamic potentials for four different cases.}
\label{tab:potentials}
\end{table}

In equations presented in Table \ref{tab:potentials}, $\mathbf{r}$ denotes radius-vector of a point, at which we consider the fields at current time $t$, relative to origin of our arbitrarily chosen reference frame. $\mathbf{r}'$ denotes radius-vector of a point, at which field is originated at retarded time $t_r$, relative to origin of our arbitrarily chosen reference frame. $\lbrcurs=\mathbf{r}-\mathbf{r}'$ denotes position of a point, at which we consider fields, relative to a point, at which the field is ``originated'' or ``emitted''. $\lrcurs$ and $\lhrcurs$ are magnitude and unit vector of $\lbrcurs$. $q$ and $\mathbf{I}$ denote electric charge and current, $\rho$ and $\mathbf{J}$ denote electric charge and current densities, while $\mathbf{v}$ is velocity of point charge or of distributed current flow. The following coordinate notation will be used for position vectors, $\mathbf{r} = x\mathbf{i}+y\mathbf{j}+z\mathbf{k}$, $\mathbf{r}' = x'\mathbf{i}+y'\mathbf{j}+z'\mathbf{k}$,  $\lbrcurs=(x-x')\mathbf{i}+(y-y')\mathbf{j}+(z-z')\mathbf{k}$, where $\mathbf{i}, \mathbf{j}, \mathbf{k}$ are Cartesian unit vector of our chosen reference frame. For all derivations in this paper we assume that our point of observation $\mathbf{r}$ is stationary relative to our reference frame, i.e.
\begin{equation}\label{eqn:global_assumption_t}
\frac{\partial{\mathbf{r}}}{\partial{t}} = \mathbf{0}
\end{equation}
Additionally in this and in all following equations nabla operator is expressed in spatial partial derivatives of coordinates associated with current time $t$ 
\begin{equation}\label{eqn:nabla}
\nabla = \frac{\partial}{\partial{x}}\mathbf{i}+\frac{\partial}{\partial{y}}\mathbf{j}+\frac{\partial}{\partial{z}}\mathbf{k}
\end{equation}
Therefore two following identities, which will be valid for all derivations in this paper, can be derived for $\mathbf{r}$
\begin{subequations}\label{eqn:global_assumption_r}
	\begin{align}
	\nabla \cdot \mathbf{r} &= \frac{\partial}{\partial{x}}x+\frac{\partial}{\partial{y}}y + \frac{\partial}{\partial{z}}z = 3 \label{subeqn01}\\
	\nabla \times \mathbf{r} &= 
	\begin{vmatrix}
	\mathbf{i} & \mathbf{j} & \mathbf{k} \\
	\frac{\partial}{\partial{x}} & \frac{\partial}{\partial{y}} & \frac{\partial}{\partial{z}} \\
	x & y & z \\
	\end{vmatrix} = 0 \label{subeqn02}
	\end{align}
\end{subequations}
Equations (\ref{eqn:global_assumption_t}), (\ref{subeqn01}) and (\ref{subeqn02}) constitute our assumptions about point $\mathbf{r}$, which can be called ``global'' assumptions, because they apply for all cases that will be considered. 

In order to derive electrodynamic fields from potentials in Table \ref{tab:potentials}, we need to consider the following 2 general conditions. First classical case will be considered. 

\subsection{Classical cases ($v \ll c$)}\label{sec:classical_case}

For a given time $t$ retarded time $t_r$ is a function of $\lrcurs$ only:
\begin{equation}\label{eqn:retarded_tr}
t_r = t - \frac{\lrcurs}{c}
\end{equation}
In other words, our current time $t$ is taken as constant, because we can arbitrarily set and/or pick its value, and so gradient of retarded time is 
\begin{equation}\label{eqn:grad_retarded_tr}
\nabla t_r = - \frac{1}{c} \nabla{\lrcurs}
\end{equation}
For gradient and time derivative of $\lrcurs$ we have
\begin{equation}\label{eqn:grad_class_retardedr}
\nabla{\lrcurs} = (\frac{\partial}{\partial{x}}\mathbf{i}+\frac{\partial}{\partial{y}}\mathbf{j}+\frac{\partial}{\partial{z}}\mathbf{k})\sqrt{(x-x')^2+(y-y')^2+(z-z')^2} = \frac{\lbrcurs}{\lrcurs} = \lhrcurs
\end{equation}
\begin{equation}\label{eqn:timed_class_retardedr}
\frac{\partial{\lrcurs}}{\partial{t}} = 0
\end{equation}
In the derivation of (\ref{eqn:grad_class_retardedr}) the assumption, that partial derivatives of $x', y', z'$ with respect to $x, y, z$ are zero, i.e. $\partial{x'}/ \partial{x} = \partial{y'} / \partial{x} = ... = \partial{y'} / \partial{z} = \partial{z'} / \partial{z} = 0$, was made. In other words, for classical case source at position $\mathbf{r}'$ moves at low speed compared to the speed of propagation of field disturbance (or information) $c$, so that in this case even for a \emph{moving} point or \emph{moving} distributed charge $\mathbf{r}'$ can be considered as almost static and constant for a given $\mathbf{r}$ and $t$. Thus electromagnetic fields at $\mathbf{r}$ and $t$ can be considered as ``emitted'' by almost ``static'' (or slowly moving compared to $c$) charges. This assumption also implies that $\partial{x'}/ \partial{t} = \partial{y'}/ \partial{t} = \partial{z'}/ \partial{t} \approx 0$, which were used to obtain (\ref{eqn:timed_class_retardedr}). In essence, as mentioned above, in classical case retardation effects can be neglected. For classical case above mentioned assumptions about $x', y', z'$ can be summarized as, in terms of $\mathbf{r}'$:
\begin{subequations}\label{eqn:local_classical_assumption_r'}
	\begin{align}
	\frac{\partial{\mathbf{r'}}}{\partial{t}} &\approx \mathbf{0} \\
	\nabla \cdot \mathbf{r'} &= \frac{\partial}{\partial{x}}x'+\frac{\partial}{\partial{y}}y' + \frac{\partial}{\partial{z}}z' = 0 \\
	\nabla \times \mathbf{r'} &= 
	\begin{vmatrix}
	\mathbf{i} & \mathbf{j} & \mathbf{k} \\
	\frac{\partial}{\partial{x}} & \frac{\partial}{\partial{y}} & \frac{\partial}{\partial{z}} \\
	x' & y' & z' \\
	\end{vmatrix} = 0 
	\end{align}
\end{subequations}
We should give a word of caution here. Original assumptions $\partial{x'}/ \partial{x} = \partial{y'} / \partial{x} = ... = \partial{y'} / \partial{z} = \partial{z'} / \partial{z} = 0$ lead to (\ref{eqn:local_classical_assumption_r'}(b, c)), but (\ref{eqn:local_classical_assumption_r'}(b, c)) don't lead to these assumptions. Let's substitute back (\ref{eqn:grad_class_retardedr}) into equation for gradient of retarded time $t_r$, (\ref{eqn:grad_retarded_tr})
\begin{equation}\label{eqn:grad_class_retardedt}
\nabla t_r = -\frac{\lhrcurs}{c}
\end{equation}
By applying assumptions in (\ref{eqn:global_assumption_t}), (\ref{eqn:global_assumption_r}), (\ref{eqn:local_classical_assumption_r'}) to the definition of $\lbrcurs$, the following identities can be found
\begin{subequations}
	\begin{align}
	\frac{\partial{\lbrcurs}}{\partial{t}} &= \mathbf{0} \label{subeqn11}\\
	\nabla \cdot \lbrcurs &= \frac{\partial}{\partial{x}}(x-x')+\frac{\partial}{\partial{y}}(y-y') + \frac{\partial}{\partial{z}}(z-z') = 3 \label{subeqn12}\\
	\nabla \times \lbrcurs &= 
	\begin{vmatrix}
	\mathbf{i} & \mathbf{j} & \mathbf{k} \\
	\frac{\partial}{\partial{x}} & \frac{\partial}{\partial{y}} & \frac{\partial}{\partial{z}} \\
	(x-x') & (y-y') & (z-z') \\
	\end{vmatrix} = 0	\label{subeqn13}
	\end{align}
\end{subequations}
If we differentiate (\ref{eqn:retarded_tr}) with respect to $t$ using (\ref{eqn:timed_class_retardedr}), then we get the identity 
\begin{equation}\label{eqn:timed_class_retardedt}
\frac{\partial{t_r}}{\partial{t}} = 1
\end{equation}
which is the same as
\begin{equation}\label{eqn:time_derivative_equality}
\frac{1}{\partial{t_r}}=\frac{1}{\partial{t}}
\end{equation}
From (\ref{eqn:grad_class_retardedr}) the following relation is obtained for gradient of $\lrcurs$ in power $n$
\begin{equation}\label{eqn:classical_grad_power_retardedr}
\nabla{\lrcurs^n} = n\lrcurs^{n-1}\nabla{\lrcurs} = n\lrcurs^{n-1}\lhrcurs
\end{equation}
where $n \in \mathbb{Z}$. For time derivative we have
\begin{equation}\label{eqn:classical_time_power_retardedr}
\frac{\partial{\lrcurs^n}}{\partial{t}} = 0
\end{equation}
Next using (\ref{subeqn11})-(\ref{subeqn13}) and (\ref{eqn:classical_grad_power_retardedr}) the following general identities can be found
\begin{subequations}
	\begin{align}
	\nabla \cdot (\lrcurs^n \lhrcurs) &= (n+2)\lrcurs^{n-1}, n \in Z, n \neq -2 \label{subeqn21} \\
	\nabla \cdot (\lrcurs^{-2} \lhrcurs) &= 4\pi \delta^3(\lbrcurs),  n=-2 \label{subeqn22} \\ 
	\nabla \times (\lrcurs^n \lhrcurs) &= 0 \label{subeqn23} \\
	\frac{\partial{(\lrcurs^n \lhrcurs)}}{\partial{t}} &= 0 \label{subeqn24}
	\end{align}
\end{subequations}
The exception for the (\ref{subeqn21}) is a case when $n=-2$, at which point (\ref{subeqn22}) becomes valid \cite{DavidGriffiths}. In (\ref{subeqn22}) $\delta^3(\lbrcurs) = \delta(x) \delta(y) \delta(z)$ is three-dimensional Dirac delta function, with the following property 
\begin{equation}
\int_{-\infty}^{\infty} \int_{-\infty}^{\infty} \int_{-\infty}^{\infty} \delta^3(\lbrcurs) dV = 1
\end{equation}
For arbitrary vector function $\mathbf{P}(\mathbf{r'},t_r)$ and a scalar function $S(\mathbf{r'},t_r)$, using (\ref{eqn:grad_class_retardedt}), the following useful identities can be derived:
\begin{equation}\label{eqn:classical_dotP}
\nabla \cdot \mathbf{P}(\mathbf{r'},t_r) = 	\frac{\partial}{\partial{x}} P_x + 	\frac{\partial}{\partial{y}} P_y + 	\frac{\partial}{\partial{z}} P_z = \nabla{t_r} \cdot \frac{\partial{\mathbf{P}}}{\partial{t_r}} = -\frac{\lhrcurs}{c} \cdot \dot{\mathbf{P}}
\end{equation}

\begin{equation}\label{eqn:classical_crossP}
\nabla \times \mathbf{P}(\mathbf{r'},t_r) = 
	\begin{vmatrix}
	\mathbf{i} & \mathbf{j} & \mathbf{k} \\
	\frac{\partial}{\partial{x}} & \frac{\partial}{\partial{y}} & \frac{\partial}{\partial{z}} \\
	P_x & P_y & P_z \\
	\end{vmatrix} = \nabla{t_r} \times \frac{\partial{\mathbf{P}}}{\partial{t_r}}  = -\frac{\lhrcurs}{c} \times \dot{\mathbf{P}}
\end{equation}

\begin{equation}\label{eqn:classical_gradS}
\nabla S(\mathbf{r'},t_r) = \frac{\partial}{\partial{x}} S \mathbf{i} + 	\frac{\partial}{\partial{y}} S \mathbf{j} + 	\frac{\partial}{\partial{z}} S \mathbf{k} = (\nabla t_r) \frac{\partial{S}}{\partial{t_r}} = -\frac{\lhrcurs}{c} \dot{S}
\end{equation}
where $\dot{\mathbf{P}} = \frac{\partial{\mathbf{P}}}{\partial{t_r}}$ notation is used, i.e. dot means differentiation with respect to retarded time $t_r$. For time derivative we have, using (\ref{eqn:time_derivative_equality})
\begin{equation}\label{eqn:classical_timeP}
\frac{\partial{\mathbf{P}}}{\partial{t}} = \mathbf{\dot{P}}
\end{equation}
\begin{equation}\label{eqn:classical_timeS}
\frac{\partial{S}}{\partial{t}} = \dot{S}
\end{equation}
Additionally the following two relations will be useful for us later
\begin{equation}\label{eqn:classical_rgradf}
	\begin{split}
(\lrcurs^n \lhrcurs \cdot \nabla) \mathbf{P}(\mathbf{r'},t_r) &= \lrcurs^{n-1}((x-x')\frac{\partial}{\partial{x}}+(y-y')\frac{\partial}{\partial{y}}+(z-z')\frac{\partial}{\partial{z}})\mathbf{P} \\ 
& = \lrcurs^{n-1}(\lbrcurs \cdot \nabla t_r) \dot{\mathbf{P}} = -\frac{\lrcurs^n}{c} \dot{\mathbf{P}}
	\end{split}
\end{equation}

\begin{equation}\label{eqn:classical_fgradr}
\begin{split}
& (\mathbf{P}(\mathbf{r'},t_r) \cdot \nabla) \lrcurs^n \lhrcurs = (P_x \frac{\partial}{\partial{x}}+P_y \frac{\partial}{\partial{y}}+P_z \frac{\partial}{\partial{z}}) \lrcurs^n \lhrcurs = \\ 
& (n-1)\lbrcurs (\mathbf{P} \cdot \nabla \lrcurs) \lrcurs^{n-2}+\lrcurs^{n-1} \mathbf{P} = (n-1) \lrcurs^{n-3} \lbrcurs (\mathbf{P} \cdot \lbrcurs) + \lrcurs^{n-1} \mathbf{P}
\end{split}
\end{equation}

\subsection{Relativistic cases ($v \sim c$)}\label{sec:relativistic_case}

Derivation of equations for the relativistic case is similar in strategy to derivation performed in classical case. However there is major difference in a fact that now we can't assume that ``source'' at $\mathbf{r'}$ and at time $t_r$ moves with low speed compared to $c$. Equations (\ref{eqn:retarded_tr}) and (\ref{eqn:grad_retarded_tr})                                                                   will be valid in relativistic case too, but there will be difference in $\nabla{\lrcurs}$:
\begin{equation}\label{eqn:grad_relat_retardedr2}
\nabla{\lrcurs} = (\frac{\partial}{\partial{x}}\mathbf{i}+\frac{\partial}{\partial{y}}\mathbf{j}+\frac{\partial}{\partial{z}}\mathbf{k})\sqrt{(x-x')^2+(y-y')^2+(z-z')^2} = \lhrcurs - (\lhrcurs \cdot \mathbf{v}) \nabla t_r
\end{equation}
During the derivation of (\ref{eqn:grad_relat_retardedr2}) the assumptions for $x', y', z'$, summarized in (\ref{eqn:local_classical_assumption_r'}), are no longer used. Instead,  we take that 
\begin{equation*}
\frac{\partial{x'}}{\partial{x}} = \frac{\partial{x'}}{\partial{t_r}} \frac{\partial{t_r}}{\partial{x}} \neq 0
\end{equation*}
and in the same way for $y'$ and $z'$, and due to these terms gradient of retarded time $t_r$ appears in (\ref{eqn:grad_relat_retardedr2}) (compare with (\ref{eqn:grad_class_retardedr})). Let's substitute (\ref{eqn:grad_relat_retardedr2}) into (\ref{eqn:grad_retarded_tr}), and after small algebraic manipulations we can find gradient of retarded time
\begin{equation}\label{eqn:grad_relat_retardedt}
\nabla t_r = -\frac{\lbrcurs}{\lrcurs c - \lbrcurs \cdot \mathbf{v}}
\end{equation}
where $\mathbf{v}$ is velocity of motion of a ``source'' at retarded position at retarded time, i.e. a ``source'' was moving with velocity $\mathbf{v}$, when it was ``emitting'' fields, which we measure later at position $\mathbf{r}$ at time $t$
\begin{equation}\label{eqn:source_velocity}
\mathbf{v} = \frac{\partial{x'}}{\partial{t_r}}\mathbf{i} + \frac{\partial{y'}}{\partial{t_r}}\mathbf{j} + \frac{\partial{z'}}{\partial{t_r}}\mathbf{k} = \mathbf{\dot{r}}'
\end{equation}
If we substitute back (\ref{eqn:grad_relat_retardedt}) into (\ref{eqn:grad_relat_retardedr2}), then we get
\begin{equation}\label{eqn:grad_relat_retardedr}
\nabla{\lrcurs} = \frac{c\lbrcurs}{\lrcurs c - \lbrcurs \cdot \mathbf{v}}
\end{equation}
If we differentiate (\ref{eqn:retarded_tr}) with respect to $t$, then we have
\begin{equation}\label{eqn:relat_trt}
\frac{\partial{t_r}}{\partial{t}} = 1 - \frac{1}{c}\frac{\partial{\lrcurs}}{\partial{t}}
\end{equation}
Next we apply differentiation with respect to $t$ to the $\lrcurs$
\begin{equation}\label{eqn:relat_rt}
\frac{\partial{\lrcurs}}{\partial{t}} = \frac{\partial{\sqrt{(x-x')^2+(y-y')^2+(z-z')^2}}}{\partial{t}} = c(1-\frac{\partial{t_r}}{\partial{t}})
\end{equation}
where the following assumptions were used for $x', y', z'$, but we write only for $x'$:
\begin{equation}
\frac{\partial{x'}}{\partial{t}} = \frac{\partial{x'}}{\partial{t_r}} \frac{\partial{t_r}}{\partial{t}} \neq 0
\end{equation}
From the (\ref{eqn:relat_trt}) and (\ref{eqn:relat_rt}) we obtain the following relations
\begin{equation}\label{eqn:timed_relat_retardedt}
\frac{\partial{t_r}}{\partial{t}} = \frac{\lrcurs c}{\lrcurs c - \lbrcurs \cdot \mathbf{v}}
\end{equation}
\begin{equation}\label{eqn:timed_relat_retardedr}
\frac{\partial{\lrcurs}}{\partial{t}} = -\frac{c(\lbrcurs \cdot \mathbf{v})}{\lrcurs c - \lbrcurs \cdot \mathbf{v}}
\end{equation}
The above mentioned assumptions about $x', y', z'$ can be summarized in terms of $\mathbf{r}'$ as following, using (\ref{eqn:grad_relat_retardedt}) and (\ref{eqn:timed_relat_retardedt}) 
\begin{subequations}\label{eqn:local_relativistic_assumption_r'}
	\begin{align}
	\frac{\partial{\mathbf{r'}}}{\partial{t}} &= \frac{\partial{x'}}{\partial{t_r}} \frac{\partial{t_r}}{\partial{t}} \mathbf{i} + \frac{\partial{y'}}{\partial{t_r}} \frac{\partial{t_r}}{\partial{t}} \mathbf{j}+ \frac{\partial{z'}}{\partial{t_r}} \frac{\partial{t_r}}{\partial{t}} \mathbf{k} = \frac{\lrcurs c \mathbf{v}}{\lrcurs c - \lbrcurs \cdot \mathbf{v}}\\
	\nabla \cdot \mathbf{r'} &= \frac{\partial{x'}}{\partial{t_r}} \frac{\partial{t_r}}{\partial{x}}+\frac{\partial{y'}}{\partial{t_r}} \frac{\partial{t_r}}{\partial{y}}+\frac{\partial{z'}}{\partial{t_r}} \frac{\partial{t_r}}{\partial{z}}=-\frac{\lbrcurs \cdot \mathbf{v}}{\lrcurs c - \lbrcurs \cdot \mathbf{v}} \\
	\nabla \times \mathbf{r'} &= 
	\begin{vmatrix}
	\mathbf{i} & \mathbf{j} & \mathbf{k} \\
	\frac{\partial}{\partial{t_r}} \frac{\partial{t_r}}{\partial{x}} & \frac{\partial}{\partial{t_r}} \frac{\partial{t_r}}{\partial{y}} & \frac{\partial}{\partial{t_r}} \frac{\partial{t_r}}{\partial{z}}\\
	x' & y' & z' \\
	\end{vmatrix} = -\frac{\lbrcurs \times \mathbf{v} }{\lrcurs c - \lbrcurs \cdot \mathbf{v}} 
	\end{align}
\end{subequations}
Using global assumptions (\ref{eqn:global_assumption_t}), (\ref{eqn:global_assumption_r}) and assumptions (\ref{eqn:local_relativistic_assumption_r'}) and the relation $\lbrcurs=\mathbf{r}-\mathbf{r}'$, the following identities are derived for $\lbrcurs$
\begin{subequations}
	\begin{align}
	\frac{\partial{\lbrcurs}}{\partial{t}} &= -\frac{\lrcurs c \mathbf{v}}{\lrcurs c - \lbrcurs \cdot \mathbf{v}} \label{subeqn31}\\
	\nabla \cdot \lbrcurs &= \frac{\partial}{\partial{x}}(x-x')+\frac{\partial}{\partial{y}}(y-y') + \frac{\partial}{\partial{z}}(z-z') = 3 + \frac{\lbrcurs \cdot \mathbf{v}}{\lrcurs c - \lbrcurs \cdot \mathbf{v}} \label{subeqn32}\\
	\nabla \times \lbrcurs &= 
	\begin{vmatrix}
	\mathbf{i} & \mathbf{j} & \mathbf{k} \\
	\frac{\partial}{\partial{x}} & \frac{\partial}{\partial{y}} & \frac{\partial}{\partial{z}} \\
	(x-x') & (y-y') & (z-z') \\
	\end{vmatrix} =  \frac{\lbrcurs \times \mathbf{v} }{\lrcurs c - \lbrcurs \cdot \mathbf{v}} 	\label{subeqn33}
	\end{align}
\end{subequations}
From (\ref{eqn:grad_relat_retardedr}) the following relation is obtained for gradient of $\lrcurs$ in power $n$
\begin{equation}\label{eqn:relativistic_grad_power_retardedr}
\nabla{\lrcurs^n} = n\lrcurs^{n-1}\nabla{\lrcurs} = n\lrcurs^{n-1} \frac{c\lbrcurs}{\lrcurs c - \lbrcurs \cdot \mathbf{v}}
\end{equation}
and using (\ref{eqn:timed_relat_retardedr}) for its time derivative
\begin{equation}\label{eqn:relativistic_time_power_retardedr}
\frac{\partial{\lrcurs^n}}{\partial{t}} = -n\lrcurs^{n-1}c\frac{\lbrcurs\cdot\mathbf{v}}{\lrcurs c - \lbrcurs \cdot \mathbf{v}}
\end{equation}
Next using (\ref{eqn:timed_relat_retardedr}), (\ref{subeqn31})-(\ref{subeqn33}), (\ref{eqn:relativistic_grad_power_retardedr}) the following general identities can be found
\begin{subequations}
	\begin{align}
	\nabla \cdot (\lrcurs^n \lhrcurs) &= \lrcurs^{n-1}(2+n\frac{\lrcurs c}{\lrcurs c - \lbrcurs \cdot \mathbf{v}}), n \in Z, n \neq -2 \label{subeqn41} \\
	\nabla \cdot (\lrcurs^{-2} \lhrcurs) &= -2\lrcurs^{-3}(\frac{\lbrcurs \cdot \mathbf{v}}{\lrcurs c - \lbrcurs \cdot \mathbf{v}})+4\pi \delta^3(\lbrcurs),  n=-2 \label{subeqn42} \\ 
	\nabla \times (\lrcurs^n \lhrcurs) &= \lrcurs^{n-1} \frac{\lbrcurs \times \mathbf{v}}{\lrcurs c - \lbrcurs \cdot \mathbf{v}} \label{subeqn43} \\
	\frac{\partial{(\lrcurs^n \lhrcurs)}}{\partial{t}} &= -(n-1)\lrcurs^{n-2}c\frac{\lbrcurs(\lbrcurs\cdot\mathbf{v})}{\lrcurs c - \lbrcurs \cdot \mathbf{v}}-\lrcurs^{n} c\frac{\mathbf{v}}{\lrcurs c - \lbrcurs \cdot \mathbf{v}} \label{subeqn44}
	\end{align}
\end{subequations}

For arbitrary vector function $\mathbf{P}(\mathbf{r'},t_r)$ and a scalar function $S(\mathbf{r'},t_r)$, using (\ref{eqn:grad_relat_retardedt}), the following identities can be derived:
\begin{equation}\label{eqn:relativistic_dotP}
\nabla \cdot \mathbf{P}(\mathbf{r'},t_r) = 	\frac{\partial{P_x}}{\partial{t_r}} \frac{\partial{t_r}}{\partial{x}} + 	\frac{\partial{P_y}}{\partial{t_r}} \frac{\partial{t_r}}{\partial{y}} + 	\frac{\partial{P_z}}{\partial{t_r}} \frac{\partial{t_r}}{\partial{z}} = \nabla{t_r} \cdot \frac{\partial{\mathbf{P}}}{\partial{t_r}} = -\frac{\lbrcurs \cdot \dot{\mathbf{P}}}{\lrcurs c - \lbrcurs \cdot \mathbf{v}}
\end{equation}

\begin{equation}\label{eqn:relativistic_crossP}
\nabla \times \mathbf{P}(\mathbf{r'},t_r) = 
\begin{vmatrix}
\mathbf{i} & \mathbf{j} & \mathbf{k} \\
\frac{\partial}{\partial{t_r}} \frac{\partial{t_r}}{\partial{x}}& \frac{\partial}{\partial{t_r}} \frac{\partial{t_r}}{\partial{y}}& \frac{\partial}{\partial{t_r}} \frac{\partial{t_r}}{\partial{z}}\\
P_x & P_y & P_z \\
\end{vmatrix} = \nabla{t_r} \times \frac{\partial{\mathbf{P}}}{\partial{t_r}}  = -\frac{\lbrcurs \times \dot{\mathbf{P}}}{\lrcurs c - \lbrcurs \cdot \mathbf{v}}
\end{equation}

\begin{equation}\label{eqn: relativistic_gradS}
\nabla S(\mathbf{r'},t_r) = \frac{\partial{S}}{\partial{t_r}} \frac{\partial{t_r}}{\partial{x}} \mathbf{i} + 	\frac{\partial{S}}{\partial{t_r}} \frac{\partial{t_r}}{\partial{y}} \mathbf{j} + 	\frac{\partial{S}}{\partial{t_r}} \frac{\partial{t_r}}{\partial{z}} \mathbf{k} = (\nabla t_r) \frac{\partial{S}}{\partial{t_r}} = -\frac{\lbrcurs}{\lrcurs c - \lbrcurs \cdot \mathbf{v}} \dot{S}
\end{equation}
where again $\dot{S} = \frac{\partial{S}}{\partial{t_r}}$ notation is used. For time derivative, using (\ref{eqn:timed_relat_retardedt}), we have 
\begin{equation}\label{eqn:relativistic_timeP}
\frac{\partial{\mathbf{P}}}{\partial{t}} = \frac{\lrcurs c}{\lrcurs c - \lbrcurs \cdot \mathbf{v}} \mathbf{\dot{P}}
\end{equation}
\begin{equation}\label{eqn:relativistic_timeS}
\frac{\partial{S}}{\partial{t}} = \frac{\lrcurs c}{\lrcurs c - \lbrcurs \cdot \mathbf{v}} \dot{S}
\end{equation}
Additionally the following three relations will be useful for us later
\begin{equation}\label{eqn:relativistic_rgradf}
\begin{split}
(\lrcurs^n \lhrcurs \cdot \nabla) \mathbf{P}(\mathbf{r'},t_r) &= \lrcurs^{n-1}((x-x')\frac{\partial}{\partial{t_r}}\frac{\partial{t_r}}{\partial{x}}+(y-y')\frac{\partial}{\partial{t_r}}\frac{\partial{t_r}}{\partial{y}}+(z-z')\frac{\partial}{\partial{t_r}}\frac{\partial{t_r}}{\partial{z}})\mathbf{P} \\ 
& = \lrcurs^{n-1}(\lbrcurs \cdot \nabla t_r) \dot{\mathbf{P}} = -\frac{\lrcurs^{n+1}}{\lrcurs c - \lbrcurs \cdot \mathbf{v}} \dot{\mathbf{P}}
\end{split}
\end{equation}

\begin{equation}\label{eqn:relativistic_fgradr}
\begin{split}
& (\mathbf{P}(\mathbf{r'},t_r) \cdot \nabla) \lrcurs^n \lhrcurs = (P_x \frac{\partial}{\partial{x}}+P_y \frac{\partial}{\partial{y}}+P_z \frac{\partial}{\partial{z}}) \lrcurs^n \lhrcurs = \\ 
& (n-1) \lrcurs^{n-3} \lbrcurs (\mathbf{P} \cdot \lbrcurs) \frac{\lrcurs c}{\lrcurs c - \lbrcurs \cdot \mathbf{v}} + \lrcurs^{n-1} (\mathbf{P} + \frac{(\mathbf{P} \cdot \lbrcurs) \mathbf{v}}{\lrcurs c - \lbrcurs \cdot \mathbf{v}})
\end{split}
\end{equation}

\begin{equation}\label{eqn:relativistic_agradb}
\begin{split}
& (\mathbf{P}(\mathbf{r'},t_r) \cdot \nabla) \mathbf{T(\mathbf{r'},t_r)} = (P_x \frac{\partial}{\partial{t_r}}\frac{\partial{t_r}}{\partial{x}}+P_y \frac{\partial}{\partial{t_r}}\frac{\partial{t_r}}{\partial{y}}+P_z \frac{\partial}{\partial{t_r}}\frac{\partial{t_r}}{\partial{z}}) \mathbf{T} = \\ 
& (\mathbf{P} \cdot \nabla t_r) \dot{\mathbf{T}} = - \frac{(\mathbf{P} \cdot \lbrcurs) \dot{\mathbf{T}}}{\lrcurs c - \lbrcurs \cdot \mathbf{v}}
\end{split}
\end{equation}
where in the last equation $\mathbf{T}(\mathbf{r'},t_r)$, like $\mathbf{P}$, is an arbitrary vector function of $\mathbf{r}'$ and $t_r$. In (\ref{eqn:relativistic_fgradr}) we should not directly apply the transformation 
\begin{equation*}
\frac{\partial}{\partial{x}} = \frac{\partial}{\partial{t_r}}\frac{\partial{t_r}}{\partial{x}}, \frac{\partial}{\partial{y}} = \frac{\partial}{\partial{t_r}}\frac{\partial{t_r}}{\partial{y}}, \frac{\partial}{\partial{z}} = \frac{\partial}{\partial{t_r}}\frac{\partial{t_r}}{\partial{z}}
\end{equation*}
as, for example, in (\ref{eqn:relativistic_agradb}). Instead we should proceed with differentiation of $\lrcurs^n \lhrcurs$ with respect to $x, y, z$ and later apply that transformation. 
The following two identities will be useful for us during the derivation
\begin{equation}\label{eqn:grad_term}
\nabla (\lrcurs c - \lbrcurs \cdot \mathbf{v}) = \frac{\lbrcurs(c^2-\mathbf{v} \cdot \mathbf{v} + \lbrcurs \cdot \mathbf{a}) - \mathbf{v}(\lrcurs c - \lbrcurs \cdot \mathbf{v})}{\lrcurs c - \lbrcurs \cdot \mathbf{v}}
\end{equation}
\begin{equation}\label{eqn:time_term}
\frac{ \partial{(\lrcurs c - \lbrcurs \cdot \mathbf{v})}}{\partial{t}} = \frac{-\lrcurs c(c^2-\mathbf{v} \cdot \mathbf{v} + \lbrcurs \cdot \mathbf{a}) + c^2(\lrcurs c - \lbrcurs \cdot \mathbf{v})}{\lrcurs c - \lbrcurs \cdot \mathbf{v}}
\end{equation}
where $\mathbf{a}$ is acceleration of a ``source'' at retarded position at retarded time
\begin{equation}\label{eqn:source_acceleration}
\mathbf{a} = \frac{\partial^2{x'}}{\partial{t_r^2}}\mathbf{i} + \frac{\partial^2{y'}}{\partial{t_r^2}}\mathbf{j} + \frac{\partial^2{z'}}{\partial{t_r^2}}\mathbf{k} = \mathbf{\dot{v}}
\end{equation}
Lastly, next two identities will be necessary for us
\begin{equation}\label{eqn:grad_term2}
\nabla (c^2-\mathbf{v} \cdot \mathbf{v} + \lbrcurs \cdot \mathbf{a}) = \frac{(3 (\mathbf{a} \cdot \mathbf{v}) - \lbrcurs \cdot \mathbf{b})\lbrcurs}{\lrcurs c - \lbrcurs \cdot \mathbf{v}} +\mathbf{a}
\end{equation}
\begin{equation}\label{eqn:time_term2}
\frac{ \partial{(c^2-\mathbf{v} \cdot \mathbf{v} + \lbrcurs \cdot \mathbf{a})}}{\partial{t}} = \frac{-\lrcurs c(3 (\mathbf{a} \cdot \mathbf{v}) - \lbrcurs \cdot \mathbf{b})}{\lrcurs c - \lbrcurs \cdot \mathbf{v}}
\end{equation}
where $\mathbf{b}$ is time derivative of acceleration, or boost, of a ``source'' at retarded position at retarded time
\begin{equation}\label{eqn:source_boost}
\mathbf{b} = \frac{\partial^3{x'}}{\partial{t_r^3}}\mathbf{i} + \frac{\partial^3{y'}}{\partial{t_r^3}}\mathbf{j} + \frac{\partial^3{z'}}{\partial{t_r^3}}\mathbf{k} = \mathbf{\dot{a}}
\end{equation}
During the derivation we often will use vector calculus identities, but in order to save space, we will not present them here. 

\section{Results}\label{sec:results}

Here we present results of our derivations. 

\subsection{Classical point charge}
\label{sec:classical_point_charge}

For a point charge we assume that its charge is constant $q=const$, i.e. doesn't change with time $t_r$ or with space coordinates $\mathbf{r}'$. This assumption is due to fundamental property of nature that a single charged particle, like electron, can not be divided (or multiplied). Potentials for this case are given in Table \ref{tab:potentials} in the upper left corner, and we present them here also
\begin{subequations}\label{eqn:classical_point_potentials}
	\begin{align}
	V(\mathbf{r},t) &= \frac{q}{4\pi \varepsilon_0 \lrcurs} \\
	\mathbf{A}(\mathbf{r},t) &= \frac{\mu_0 \mathbf{I}}{4\pi \lrcurs}
	\end{align}
\end{subequations}
Current and charge are related to each other $\mathbf{I} = q\mathbf{v} = \emph{const}$. Here we should stop and think about meaning of vector potential $\mathbf{A}$. From rigorous point of view, a moving point charge can't generate steady current \cite{DavidGriffiths}. Therefore current $\mathbf{I}$ should be called pseudo-current, but for simplicity we will continue to call it current. For classical case we assume that charge is moving slowly, i.e. $\mathbf{v} \approx 0$ and (\ref{eqn:local_classical_assumption_r'}). Therefore even from strict mathematical point of view we need to take $\mathbf{I}=0$ and $\mathbf{A}=0$ when $\mathbf{v} \approx 0$, as long as $\mathbf{v} \ll c$, we can assume that there is small nonzero current, which gives rise to nonzero $\mathbf{A}$. Due to the fact that we can ignore retardation effects, we assume that current is constant $\mathbf{I}=\emph{const}$. In other words, motion of a charge, at the moment when it ``emits'' the fields, does not affect the fields. Thus current does not depend on $t, t_r, \mathbf{r}, \mathbf{r}'$. Using (\ref{eqn:classical_grad_power_retardedr}) and (\ref{subeqn22}), we can find gradient and Laplacian of $V$
\begin{subequations}
	\begin{align}
	\nabla V(\mathbf{r},t) &= \frac{q}{4\pi \varepsilon_0} \nabla(\frac{1}{\lrcurs}) = -\frac{q}{4\pi \varepsilon_0} \frac{\lhrcurs}{\lrcurs^2} \label{eqn:cp_nablaV}\\
	\nabla^2 V(\mathbf{r},t) &= -\frac{q}{4\pi \varepsilon_0} \nabla \cdot (\frac{\lhrcurs}{\lrcurs^2}) = -\frac{q}{\varepsilon_0} \delta^3(\lbrcurs) \label{eqn:cp_nabla2V}
	\end{align}
\end{subequations}
Using (\ref{eqn:timed_class_retardedr}) the following two identities are found:
\begin{subequations}
	\begin{align}
	\frac{\partial{V(\mathbf{r},t)} }{\partial{t}} &= \frac{q}{4\pi \varepsilon_0} \frac{\partial{} }{\partial{t}}(\frac{1}{\lrcurs}) = 0 \label{eqn:cp_timeV}\\
	\frac{\partial^2{V(\mathbf{r},t)}}{\partial{t^2}} &= 0 \label{eqn:cp_time2V}
	\end{align}
\end{subequations}
Substituting (\ref{eqn:cp_nabla2V}) and (\ref{eqn:cp_time2V}) into (\ref{eqn:Maxwell_Lorentz}), we get
\begin{equation}
\square^2 V = -\frac{q}{\varepsilon_{0}} \delta^3(\lbrcurs) \label{eqn:cp_V_Lorentz}
\end{equation}
Thus Lorentz gauge assumption in terms of scalar potential $V$ is satisfied. Now we do the same analysis for vector potential $\mathbf{A}(\mathbf{r},t)$ using (\ref{eqn:classical_grad_power_retardedr}), (\ref{subeqn22}), (\ref{subeqn23}) and some vector calculus identities. 
\begin{subequations}
	\begin{align}
	\nabla \cdot \mathbf{A} &= \frac{\mu_0 \mathbf{I}}{4 \pi} \cdot \nabla{(\frac{1}{\lrcurs})} = -\frac{\mu_0 \mathbf{I}\cdot\lhrcurs}{4 \pi \lrcurs^2} \label{eqn:cp_nabladotA} \\
	\nabla(\nabla \cdot \mathbf{A}) &= -\frac{\mu_0}{4 \pi} \nabla(\frac{\mathbf{I} \cdot \lhrcurs}{\lrcurs^2}) = \frac{\mu_0}{4 \pi} (3\frac{\lhrcurs(\mathbf{I}\cdot \lhrcurs)}{\lrcurs^3}-\frac{\mathbf{I}}{\lrcurs^3}) \label{eqn:cp_nabladot2A} \\
	\nabla \times \mathbf{A} &= -\frac{\mu_0 \mathbf{I}}{4 \pi} \times \nabla{(\frac{1}{\lrcurs})} = \frac{\mu_0 \mathbf{I}\times\lhrcurs}{4 \pi \lrcurs^2} \label{eqn:cp_nablacrossA} \\
	\nabla \times (\nabla \times \mathbf{A}) &= \frac{\mu_0}{4 \pi} \nabla \times (\frac{\mathbf{I} \times \lhrcurs}{\lrcurs^2}) = \frac{\mu_0}{4 \pi} (3\frac{\lhrcurs(\mathbf{I}\cdot \lhrcurs)}{\lrcurs^3}-\frac{\mathbf{I}}{\lrcurs^3}+4 \pi \mathbf{I} \delta^3(\lbrcurs)) \label{eqn:cp_nablacross2A}
	\end{align}
\end{subequations}
Now we find time derivatives of $\mathbf{A}(\mathbf{r},t)$ using (\ref{eqn:timed_class_retardedr})
\begin{subequations}
	\begin{align}
	\frac{\partial{\mathbf{A}(\mathbf{r},t)} }{\partial{t}} &= \frac{\mu_0 \mathbf{I}}{4\pi} \frac{\partial{} }{\partial{t}}(\frac{1}{\lrcurs}) = 0 \label{eqn:cp_timeA}\\
	\frac{\partial^2{\mathbf{A}(\mathbf{r},t)}}{\partial t^2} &= 0 \label{eqn:cp_time2A}
	\end{align}
\end{subequations}
Substituting back (\ref{eqn:cp_nabladot2A}), (\ref{eqn:cp_nablacross2A}) and (\ref{eqn:cp_time2A}) into (\ref{eqn:Maxwell_Lorentz}), we get
\begin{equation}
\square^2 \mathbf{A} = -\mu_0 \mathbf{I} \delta^3(\lbrcurs) \label{eqn:cp_A_Lorentz}
\end{equation}
Therefore Lorentz gauge condition for $\mathbf{A}$ is satisfied. Let's check if Lorentz gauge function is zero by substituting (\ref{eqn:cp_timeV}) and (\ref{eqn:cp_nabladotA}) into (\ref{eqn:Lorentz_gauge}).
\begin{equation}
L = \nabla \cdot \mathbf{A} + \mu_0 \varepsilon_{0} \frac{\partial{V}}{\partial{t}} = -\frac{\mu_0 \mathbf{I}\cdot\lhrcurs}{4 \pi \lrcurs^2} \label{eqn:cp_Lorentz}
\end{equation} 
We see that strictly speaking, Lorentz gauge condition is not satisfied, but if we take exact cases when $\mathbf{I}=0$ or when $\mathbf{I}$ is perpendicular to $\lhrcurs$, then it is satisfied. 
Let's now find fields. By substituting (\ref{eqn:cp_nablaV}), (\ref{eqn:cp_nablacrossA}), (\ref{eqn:cp_timeA}) into (\ref{eqn:vector_potential}) and (\ref{eqn:scalar_potential}) we get
\begin{subequations}
	\begin{align}
		\mathbf{E} &= \frac{q}{4\pi\varepsilon_0}\frac{\lhrcurs}{\lrcurs^2} \label{eqn:cp_E}\\
		\mathbf{B} &= \frac{\mu_0 \mathbf{I}\times\lhrcurs}{4 \pi \lrcurs^2} \label{eqn:cp_B}
	\end{align}
\end{subequations}
These are electrostatic and magnetostatic fields of a point charge. As mentioned in \cite{DavidGriffiths}, (\ref{eqn:cp_B}) is approximately correct for a point charge when we can neglect with retardation, which is precisely the case we are considering here. Finally, in order to see if Gauss's law is satisfied for these electric and magnetic fields, let's find divergence of fields by using (\ref{eqn:classical_grad_power_retardedr}), (\ref{subeqn22}), (\ref{subeqn23}), (\ref{eqn:cp_E}), (\ref{eqn:cp_B}) 
\begin{subequations}
	\begin{align}
	\nabla \cdot \mathbf{E} &= \frac{q}{4\pi\varepsilon_0} \nabla \cdot (\frac{\lhrcurs}{\lrcurs^2}) = \frac{q}{\varepsilon_0} \delta^3(\lbrcurs) \label{eqn:cp_nablaE} \\
	\nabla \cdot \mathbf{B} &= \frac{\mu_0}{4\pi} \nabla \cdot (\frac{\mathbf{I} \times \lhrcurs}{\lrcurs^2}) = 0 \label{eqn:cp_nablaB} 
	\end{align}
\end{subequations}
These results give us back original equations from electrostatics and magnetostatics. In essence, this case corresponds to static electric and magnetic fields of a point charge. For example, all time derivatives of potentials are zero. Gauss's law is satisfied, nothing is new in this case.

\subsection{Relativistic point charge} \label{sec:relativistic_point_charge}

For a relativistic point charge we keep the assumption that charge is constant. The electrodynamic potentials, called Li\'enard-Wiechert potentials, are given in the upper right corner of the Table \ref{tab:potentials}, and we present them here also
\begin{subequations}\label{eqn:relativistic_point_potentials}
	\begin{align}
	V(\mathbf{r},t) &= \frac{qc}{4\pi \varepsilon_0 (\lrcurs c - \lbrcurs \cdot \mathbf{v})}  \\
	\mathbf{A}(\mathbf{r},t) &= \frac{\mu_0 qc \mathbf{v}}{4\pi (\lrcurs c - \lbrcurs \cdot \mathbf{v})}
	\end{align}
\end{subequations}
There is slight difference in variables used in vector potential $\mathbf{A}$ given in (\ref{eqn:classical_point_potentials}) and in (\ref{eqn:relativistic_point_potentials}), in that in the former there is current $\mathbf{I}$, while in the latter there is velocity $\mathbf{v}$, and $\mathbf{I} = q\mathbf{v}$. 

Taking into account (\ref{eqn:relativistic_dotP}), (\ref{eqn:grad_term}) and (\ref{eqn:grad_term2}), we find for scalar potential
\begin{subequations}
	\begin{align}
	\nabla V(\mathbf{r},t) &= \frac{qc}{4\pi \varepsilon_0} \nabla\Big(\frac{1}{\lrcurs c - \lbrcurs \cdot \mathbf{v}}\Big) \nonumber \\ 
	& = -\frac{qc}{4\pi \varepsilon_0} \frac{\lbrcurs(c^2-\mathbf{v} \cdot \mathbf{v} + \lbrcurs \cdot \mathbf{a}) - \mathbf{v}(\lrcurs c - \lbrcurs \cdot \mathbf{v})}{(\lrcurs c - \lbrcurs \cdot \mathbf{v})^3} \label{eqn:rp_nablaV} \\
	\nabla^2 V(\mathbf{r},t) &= -\frac{qc}{4\pi \varepsilon_0} \Big(\frac{1}{(\lrcurs c - \lbrcurs \cdot \mathbf{v})^3}(3c^2-5\mathbf{v} \cdot \mathbf{v} + 5 \lbrcurs \cdot \mathbf{a}) \nonumber \\
	+ & \frac{1}{(\lrcurs c - \lbrcurs \cdot \mathbf{v})^4}((\lbrcurs \cdot \lbrcurs)(3(\mathbf{v} \cdot \mathbf{a})-\lbrcurs \cdot \mathbf{b}) + 6(c^2-\mathbf{v} \cdot \mathbf{v} + \lbrcurs \cdot \mathbf{a})(\lbrcurs \cdot \mathbf{v})) \nonumber \\
	+ & \frac{1}{(\lrcurs c - \lbrcurs \cdot \mathbf{v})^5} (-3(c^2-\mathbf{v} \cdot \mathbf{v} + \lbrcurs \cdot \mathbf{a})^2 (\lbrcurs \cdot \lbrcurs)) \Big) \label{eqn:rp_nabla2V}
	\end{align}
\end{subequations}
Using (\ref{eqn:timed_relat_retardedr}), (\ref{eqn:time_term}) and (\ref{eqn:time_term2}), and after some algebraic manipulation, we can find time derivatives of scalar potential
\begin{subequations}
	\begin{align}
	\frac{\partial{V(\mathbf{r},t)}}{\partial{t}} &= \frac{qc}{4\pi \varepsilon_0}\frac{\partial{}}{\partial{t}} \Big(\frac{1}{\lrcurs c - \lbrcurs \cdot \mathbf{v}}\Big) \nonumber \\
	& = \frac{qc^2}{4\pi \varepsilon_0}\frac{(\lrcurs(c^2 - \mathbf{v} \cdot \mathbf{v} + \lbrcurs \cdot \mathbf{a}) -c(\lrcurs c - \lbrcurs \cdot \mathbf{v}))}{(\lrcurs c - \lbrcurs \cdot \mathbf{v})^3} \label{eqn:rp_timeV} \\
	\frac{\partial^2{V(\mathbf{r},t)}}{\partial{t^2}} &= -\frac{qc^3}{4\pi \varepsilon_0} \Big( \frac{1}{(\lrcurs c - \lbrcurs \cdot \mathbf{v})^3} (3c^2-5\mathbf{v} \cdot \mathbf{v} + 5 \lbrcurs \cdot \mathbf{a}) \nonumber \\
	+ & \frac{1}{(\lrcurs c - \lbrcurs \cdot \mathbf{v})^4}((\lbrcurs \cdot \lbrcurs)(3(\mathbf{v} \cdot \mathbf{a})-\lbrcurs \cdot \mathbf{b}) + 6(c^2-\mathbf{v} \cdot \mathbf{v} + \lbrcurs \cdot \mathbf{a})(\lbrcurs \cdot \mathbf{v})) \nonumber \\
	+ & \frac{1}{(\lrcurs c - \lbrcurs \cdot \mathbf{v})^5} (-3(c^2-\mathbf{v} \cdot \mathbf{v} + \lbrcurs \cdot \mathbf{a})^2 (\lbrcurs \cdot \lbrcurs)) \Big) \label{eqn:rp_time2V} 
	\end{align}
\end{subequations}
If we substitute (\ref{eqn:rp_nabla2V}) and (\ref{eqn:rp_time2V}) into (\ref{eqn:Maxwell_Lorentz}), then we get
\begin{equation}
\square^2 V = 0 \label{eqn:rp_V_Lorentz}
\end{equation}
This result shows that Lorentz gauge condition for $V$ in this case has trivial solution. 

Using (\ref{eqn:relativistic_dotP}), (\ref{eqn:relativistic_crossP}), (\ref{eqn:relativistic_rgradf})-(\ref{eqn:relativistic_agradb}), (\ref{eqn:grad_term}), (\ref{eqn:grad_term2}) and some vector calculus identities, we find for vector potential $\mathbf{A}$
\begin{subequations}
	\begin{align}
	\nabla \cdot \mathbf{A} &= \frac{\mu_0 qc}{4 \pi} \nabla \cdot \Big(\frac{\mathbf{v}}{\lrcurs c - \lbrcurs \cdot \mathbf{v}}\Big) \nonumber \\
	& = -\frac{\mu_0 qc}{4 \pi}\frac{\Big((\lbrcurs \cdot \mathbf{v})(c^2-\mathbf{v} \cdot \mathbf{v} + \lbrcurs \cdot \mathbf{a})+(\lbrcurs \cdot \mathbf{a} - \mathbf{v} \cdot \mathbf{v})(\lrcurs c - \lbrcurs \cdot \mathbf{v})\Big)}{(\lrcurs c - \lbrcurs \cdot \mathbf{v})^3} \label{eqn:rp_nabladotA} \\	
	\nabla(\nabla \cdot \mathbf{A}) &= -\frac{\mu_0 qc}{4 \pi} \Big(\frac{1}{(\lrcurs c - \lbrcurs \cdot \mathbf{v})^3} (\lrcurs c \mathbf{a} - 2c^2 \mathbf{v}) \nonumber \\
	& + \frac{1}{(\lrcurs c - \lbrcurs \cdot \mathbf{v})^4} (3(\lrcurs c \mathbf{v} -\lbrcurs(-\mathbf{v} \cdot \mathbf{v}+\lbrcurs \cdot \mathbf{a}))(c^2-\mathbf{v} \cdot \mathbf{v} + \lbrcurs \cdot \mathbf{a}) \nonumber \\
	& + \lrcurs c \lbrcurs (3(\mathbf{v} \cdot \mathbf{a}) - \lbrcurs \cdot \mathbf{b})) +  \frac{1}{(\lrcurs c - \lbrcurs \cdot \mathbf{v})^5} (-3\lbrcurs (\lbrcurs \cdot \mathbf{v}) (c^2-\mathbf{v} \cdot \mathbf{v} + \lbrcurs \cdot \mathbf{a})^2) \Big) \label{eqn:rp_nabladot2A} \\
	\nabla \times \mathbf{A} &= \frac{\mu_0 qc}{4 \pi} \nabla \times \Big(\frac{\mathbf{v}}{\lrcurs c - \lbrcurs \cdot \mathbf{v}}\Big) \nonumber \\
	& = -\frac{\mu_0 qc}{4 \pi}\frac{\Big((\lbrcurs \times \mathbf{v})(c^2-\mathbf{v} \cdot \mathbf{v} + \lbrcurs \cdot \mathbf{a})+(\lbrcurs \times \mathbf{a})(\lrcurs c - \lbrcurs \cdot \mathbf{v})\Big)}{(\lrcurs c - \lbrcurs \cdot \mathbf{v})^3} \label{eqn:rp_nablacrossA} \\
	\nabla \times (\nabla \times \mathbf{A}) &= -\frac{\mu_0 qc}{4 \pi} \Big(\frac{1}{(\lrcurs c - \lbrcurs \cdot \mathbf{v})^3} ((-2\lrcurs c \mathbf{a} - (\lbrcurs \cdot \mathbf{v}) \mathbf{a} + \lrcurs^2 \mathbf{b}) + \mathbf{v}(c^2-\mathbf{v} \cdot \mathbf{v} + \lbrcurs \cdot \mathbf{a})) \nonumber \\
	& + \frac{1}{(\lrcurs c - \lbrcurs \cdot \mathbf{v})^4}(3(-\lrcurs c \mathbf{v} - \lbrcurs(-\mathbf{v} \cdot \mathbf{v}+\lbrcurs \cdot \mathbf{a})+(\lbrcurs \cdot \lbrcurs) \mathbf{a})(c^2-\mathbf{v} \cdot \mathbf{v} + \lbrcurs \cdot \mathbf{a}) \nonumber \\
	& + (\lrcurs c \lbrcurs - (\lbrcurs \cdot \lbrcurs)\mathbf{v}) (3(\mathbf{v} \cdot \mathbf{a}) - \lbrcurs \cdot \mathbf{b})) \nonumber \\
	& + \frac{1}{(\lrcurs c - \lbrcurs \cdot \mathbf{v})^5} (-3(\lbrcurs (\lbrcurs \cdot \mathbf{v})-\mathbf{v}(\lbrcurs \cdot \lbrcurs)) (c^2-\mathbf{v} \cdot \mathbf{v} + \lbrcurs \cdot \mathbf{a})^2) \Big) \label{eqn:rp_nablacross2A}
	\end{align}
\end{subequations}
Now we need to compute time derivative of $\mathbf{A}$, using (\ref{eqn:timed_relat_retardedr}), (\ref{subeqn44}), (\ref{eqn:relativistic_timeP}), (\ref{eqn:relativistic_timeS}), (\ref{eqn:time_term}) and (\ref{eqn:time_term2}), we get
\begin{subequations}
	\begin{align}
	\frac{\partial{\mathbf{A}(\mathbf{r},t)} }{\partial{t}} &= \frac{\mu_0 qc}{4 \pi} \frac{\partial}{\partial{t}} \Big(\frac{\mathbf{v}}{\lrcurs c - \lbrcurs \cdot \mathbf{v}}\Big) \nonumber \\
	& = -\frac{\mu_0 qc^2}{4 \pi}\frac{\Big(-\lrcurs \mathbf{v}(c^2-\mathbf{v} \cdot \mathbf{v} + \lbrcurs \cdot \mathbf{a})+(c\mathbf{v} - \lrcurs \mathbf{a})(\lrcurs c - \lbrcurs \cdot \mathbf{v})\Big)} {(\lrcurs c - \lbrcurs \cdot \mathbf{v})^3} \label{eqn:rp_timeA}\\
	\frac{\partial^2{\mathbf{A}(\mathbf{r},t)}}{\partial t^2} &= -\frac{\mu_0 qc^3}{4 \pi} \Big(\frac{1}{(\lrcurs c - \lbrcurs \cdot \mathbf{v})^3} ((3\lrcurs c \mathbf{a} + (\lbrcurs \cdot \mathbf{v}) \mathbf{a} - 2c^2 \mathbf{v} - \lrcurs^2 \mathbf{b})-\mathbf{v}(c^2-\mathbf{v} \cdot \mathbf{v} + \lbrcurs \cdot \mathbf{a})) \nonumber \\
	& + \frac{1}{(\lrcurs c - \lbrcurs \cdot \mathbf{v})^4} (3(2\lrcurs c \mathbf{v} -(\lbrcurs \cdot \lbrcurs) \mathbf{a})(c^2-\mathbf{v} \cdot \mathbf{v} + \lbrcurs \cdot \mathbf{a}) \nonumber \\
	& + (\lbrcurs \cdot \lbrcurs)\mathbf{v} (3(\mathbf{v} \cdot \mathbf{a}) - \lbrcurs \cdot \mathbf{b})) +\frac{1}{(\lrcurs c - \lbrcurs \cdot \mathbf{v})^5} (-3(\mathbf{v} (\lbrcurs \cdot \lbrcurs)) (c^2-\mathbf{v} \cdot \mathbf{v} + \lbrcurs \cdot \mathbf{a})^2) \Big) \label{eqn:rp_time2A}
	\end{align}
\end{subequations}
If we substitute (\ref{eqn:rp_nabladot2A}), (\ref{eqn:rp_nablacross2A}) and (\ref{eqn:rp_time2A}) into (\ref{eqn:Maxwell_Lorentz}), then we get
\begin{equation}
\square^2 \mathbf{A} = 0 \label{eqn:rp_A_Lorentz}
\end{equation}
Which again shows that Lorentz gauge condition for $\mathbf{A}$ assumes trivial solution for a relativistic point charge.  Let's now use (\ref{eqn:rp_timeV}) and (\ref{eqn:rp_nabladotA}) to find Lorentz gauge function in (\ref{eqn:Lorentz_gauge}):
\begin{equation}
L = \nabla \cdot \mathbf{A} + \mu_0 \varepsilon_{0} \frac{\partial{V}}{\partial{t}} = 0 \label{eqn:rp_Lorentz}
\end{equation} 
which tells us that relativistic point charge satisfies Lorentz gauge condition exactly. Equations (\ref{eqn:rp_nablaV}), (\ref{eqn:rp_nablacrossA}) and (\ref{eqn:rp_timeA}) are used in order to find fields
\begin{subequations}
	\begin{align}
	\mathbf{E} &= \frac{q}{4\pi\varepsilon_0}\Big(\frac{1}{(\lrcurs c - \lbrcurs \cdot \mathbf{v})^2}(-(c\mathbf{v}+\lrcurs \mathbf{a})) \nonumber \\
	& + \frac{1}{(\lrcurs c - \lbrcurs \cdot \mathbf{v})^3}(\lbrcurs c(c^2-\mathbf{v} \cdot \mathbf{v} + \lbrcurs \cdot \mathbf{a})+\mathbf{v}(-c \lbrcurs \cdot \mathbf{v} + \lrcurs (\mathbf{v} \cdot \mathbf{v}-\lbrcurs \cdot \mathbf{a})))\Big)\label{eqn:rp_E}\\
	\mathbf{B} &= -\frac{\mu_0 qc}{4 \pi}\frac{\Big((\lbrcurs \times \mathbf{v})(c^2-\mathbf{v} \cdot \mathbf{v} + \lbrcurs \cdot \mathbf{a})+(\lbrcurs \times \mathbf{a})(\lrcurs c - \lbrcurs \cdot \mathbf{v})\Big)}{(\lrcurs c - \lbrcurs \cdot \mathbf{v})^3} \label{eqn:rp_B}
	\end{align}
\end{subequations}
Equation (\ref{eqn:rp_B}) is the same as (\ref{eqn:rp_nablacrossA}). If we introduce variable $\mathbf{u} = c\lhrcurs - \mathbf{v}$, then equations for $\mathbf{E}$ and $\mathbf{B}$ can be re-written in more familiar form as
\begin{subequations}
	\begin{align}
	\mathbf{E} &= \frac{q}{4\pi\varepsilon_0}\frac{\lrcurs}{(\lbrcurs \cdot \mathbf{u})^3} \Big(\mathbf{u}(c^2-\mathbf{v} \cdot \mathbf{v})+\lbrcurs \times (\mathbf{u} \times \mathbf{a})\Big)\\
	\mathbf{B} &= -\frac{\mu_0 qc}{4 \pi}\frac{\Big((\lbrcurs \times \mathbf{v})(c^2-\mathbf{v} \cdot \mathbf{v} + \lbrcurs \cdot \mathbf{a})+(\lbrcurs \times \mathbf{a})(\lbrcurs \cdot \mathbf{u})\Big)}{(\lbrcurs \cdot \mathbf{u})^3}
	\end{align}
\end{subequations}
Next divergence of electric and magnetic fields is found using (\ref{eqn:grad_term}), (\ref{eqn:grad_term2}), (\ref{eqn:rp_E}), (\ref{eqn:rp_B}) and some vector calculus identities
\begin{subequations}
	\begin{align}
	\nabla \cdot \mathbf{E} &= 0 \label{eqn:rp_nablaE} \\
	\nabla \cdot \mathbf{B} &= 0 \label{eqn:rp_nablaB} 
	\end{align}
\end{subequations}													
As it can be observed, in Lorentz potential divergence of electric and of magnetic field is exactly zero no matter how a charge moves. Again Gauss's law is satisfied, nothing is new in this case. 
					
\subsection{Classical continuous charge} \label{sec:classical_distributed_charge}

In this section we focus our attention on distributed continuous charge. For continuous charge we assume that charge density and current density are functions of retarded time $t_r$ and of space $\mathbf{r}'$: $\rho = \rho(\mathbf{r}',t_r)$ and $\mathbf{J} = \mathbf{J}(\mathbf{r}',t_r)$. In other words, the difference between a point and distributed charges is that for the latter we can no longer take the assumption that charge (or charge density) is constant. In other words, charge and current densities are assumed to vary in time and in space, as for example, density and velocity of compressible fluid. Electric and magnetic potentials for this case are given in Table \ref{tab:potentials} in the lower left corner
\begin{subequations}\label{eqn:classical_distributed_potentials}
	\begin{align}
		V(\mathbf{r},t) = \frac{1}{4\pi \varepsilon_0} \int \frac{\rho(\mathbf{r}',t_r)}{\lrcurs} d\tau' \\
		\mathbf{A}(\mathbf{r},t) = \frac{\mu_0}{4\pi} \int \frac{\mathbf{J}(\mathbf{r}',t_r)}{\lrcurs} d\tau'
	\end{align}
\end{subequations}
where $d\tau' = dx'dy'dz'$ is infinitesimal volume of space, defined by $\mathbf{r}'$. For distributed charges we will often use Leibniz integral rule to interchange the order of differentiation with integration. Therefore we assume that boundary of connected region, where integration by $d\tau'$ takes place, is fixed relative to global $xyz$ frame and so does not move. This is the simplest case, in which terms in Leibniz integral rule related to boundary motion vanish. In other words, integration with respect to $\tau'$ is decoupled and is independent from differentiation with respect to $t$ or $x$ or $y$ or $z$. Thus using (\ref{eqn:classical_grad_power_retardedr}), (\ref{subeqn21}), (\ref{subeqn22}), (\ref{eqn:classical_gradS}) and some vector calculus identities, we can find divergence and Laplacian of scalar potential
\begin{subequations}
	\begin{align}
	\nabla V(\mathbf{r},t) &= \frac{1}{4\pi \varepsilon_0} \int \nabla(\frac{\rho}{\lrcurs}) d\tau' = -\frac{1}{4\pi \varepsilon_0} \int (\frac{\dot{\rho}}{\lrcurs c} \lhrcurs + \frac{\rho}{\lrcurs^2}\lhrcurs) d\tau' \label{eqn:cd_nablaV}\\
	\nabla^2 V(\mathbf{r},t) &= \frac{1}{4\pi \varepsilon_0} \int (\frac{\ddot{\rho}}{\lrcurs c^2}  - \rho 4 \pi \delta^3(\lbrcurs)) d\tau'  \label{eqn:cd_nabla2V}
	\end{align}
\end{subequations}
Using (\ref{eqn:timed_class_retardedr}) time derivatives of scalar potential are found
\begin{subequations}
	\begin{align}
	\frac{\partial{V(\mathbf{r},t)} }{\partial{t}} &= \frac{1}{4\pi \varepsilon_0} \int \frac{\partial}{\partial{t}} (\frac{\rho}{\lrcurs}) d\tau' = \frac{1}{4\pi \varepsilon_0} \int \frac{\dot{\rho}}{\lrcurs} d\tau' \label{eqn:cd_timeV}\\
	\frac{\partial^2{V(\mathbf{r},t)}}{\partial{t^2}} &= \frac{1}{4\pi \varepsilon_0} \int \frac{\ddot{\rho}}{\lrcurs} d\tau' \label{eqn:cd_time2V}
	\end{align}
\end{subequations}
Let's substitute (\ref{eqn:cd_nabla2V}) and (\ref{eqn:cd_time2V}) into (\ref{eqn:Maxwell_Lorentz})
\begin{equation}
\square^2 V = - \frac{1}{\varepsilon_0} \int \rho(\mathbf{r}',t_r) \delta^3(\lbrcurs) d\tau' = -\frac{\rho(\mathbf{r},t)}{\varepsilon_{0}} \label{eqn:cd_V_Lorentz}
\end{equation}
where Dirac delta function will pick up those values of charge density, where $\mathbf{r}' = \mathbf{r}$ and $t_r = t$, i.e. $\rho = \rho(\mathbf{r},t)$. As we can see Lorentz gauge condition for $V$ is satisfied. Now let's focus on vector potential $\mathbf{A}$. 
\begin{subequations}
	\begin{align}
	\nabla \cdot \mathbf{A} &= \frac{\mu_0}{4 \pi} \int \nabla \cdot (\frac{\mathbf{J}}{\lrcurs}) d\tau' = - \frac{\mu_0}{4 \pi} \int (\frac{\mathbf{\dot{J}} \cdot \lhrcurs}{\lrcurs c} + \frac{\mathbf{J} \cdot \lhrcurs}{\lrcurs^2}) d\tau' \label{eqn:cd_nabladotA} \\
	\nabla(\nabla \cdot \mathbf{A}) &= \frac{\mu_0}{4 \pi} \int \Big( \frac{\lhrcurs}{\lrcurs c} (\frac{\lhrcurs}{c} \cdot \mathbf{\ddot{J}}) - \mathbf{\ddot{J}}(\frac{\lhrcurs}{\lrcurs c} \cdot \frac{\lhrcurs}{c})+ \frac{\mathbf{\ddot{J}}}{\lrcurs c^2} + 2\frac{\lhrcurs}{\lrcurs^2 c}(\mathbf{\dot{J}} \cdot \lhrcurs) + \nonumber \\ 
	& \frac{\lhrcurs}{c}(\frac{\lhrcurs}{\lrcurs^2} \cdot \mathbf{\dot{J}}) - \mathbf{\dot{J}}(\frac{\lhrcurs}{\lrcurs^2} \cdot \frac{\lhrcurs}{c}) + 3\frac{\lhrcurs}{\lrcurs^3}(\mathbf{J} \cdot \lhrcurs) - \frac{\mathbf{J}}{\lrcurs^3} \Big) d\tau' \label{eqn:cd_nabladot2A} \\
	\nabla \times \mathbf{A} &= \frac{\mu_0}{4 \pi} \int \nabla \times (\frac{\mathbf{J}}{\lrcurs}) d\tau' = \frac{\mu_0}{4 \pi} \int (\frac{\mathbf{\dot{J}} \times \lhrcurs}{\lrcurs c} + \frac{\mathbf{J} \times \lhrcurs}{\lrcurs^2}) d\tau' \label{eqn:cd_nablacrossA} \\
	\nabla \times (\nabla \times \mathbf{A}) &= \frac{\mu_0}{4 \pi} \int \Big( \frac{\lhrcurs}{\lrcurs c} (\frac{\lhrcurs}{c} \cdot \mathbf{\ddot{J}}) - \frac{\mathbf{\ddot{J}}}{\lrcurs c^2} + 3\frac{\lhrcurs}{\lrcurs^2 c}(\mathbf{\dot{J}} \cdot \lhrcurs) - \nonumber \\ 
	&  \mathbf{\dot{J}}(\frac{\lhrcurs}{\lrcurs^2} \cdot \frac{\lhrcurs}{c}) + 3\frac{\lhrcurs}{\lrcurs^3}(\mathbf{J} \cdot \lhrcurs) - \frac{\mathbf{J}}{\lrcurs^3} + 4 \pi \delta^3(\lbrcurs) \mathbf{J} \Big) d\tau' \label{eqn:cd_nablacross2A}
	\end{align}
\end{subequations}
Now again using (\ref{eqn:timed_class_retardedr}) we can find time derivatives of vector potential
\begin{subequations}
	\begin{align}
	\frac{\partial{\mathbf{A}(\mathbf{r},t)} }{\partial{t}} &= \frac{\mu_0}{4 \pi} \int \frac{\partial}{\partial{t}}(\frac{\mathbf{J}}{\lrcurs}) d\tau' = \frac{\mu_0}{4 \pi} \int (\frac{\mathbf{\dot{J}}}{\lrcurs}) d\tau' \label{eqn:cd_timeA}\\
	\frac{\partial^2{\mathbf{A}(\mathbf{r},t)}}{\partial t^2} &= \frac{\mu_0}{4 \pi} \int (\frac{\mathbf{\ddot{J}}}{\lrcurs}) d\tau' \label{eqn:cd_time2A}
	\end{align}
\end{subequations}
If we substitute (\ref{eqn:cd_nabladot2A}), (\ref{eqn:cd_nablacross2A})
and (\ref{eqn:cd_time2A}) into (\ref{eqn:Maxwell_Lorentz}), then 
\begin{equation}
\square^2 \mathbf{A} = -\mu_0 \int \mathbf{J}(\mathbf{r}',t_r) \delta^3(\lbrcurs) d\tau' = -\mu_0 \mathbf{J}(\mathbf{r},t) \label{eqn:cd_A_Lorentz}
\end{equation}
where again Dirac delta function will pick up those values of current density, where $\mathbf{r}' = \mathbf{r}$ and $t_r = t$, i.e. $\mathbf{J} = \mathbf{J}(\mathbf{r},t)$. Here we again notice that Lorentz gauge condition is satisfied for vector potential $\mathbf{A}$. 
Let's now check the value of the Lorentz gauge function (\ref{eqn:Lorentz_gauge}), using (\ref{eqn:cd_timeV}) and (\ref{eqn:cd_nabladotA})
\begin{equation}
	L = \nabla \cdot \mathbf{A} + \mu_0 \varepsilon_{0} \frac{\partial{V}}{\partial{t}} = -\frac{\mu_0}{4 \pi } \int  \frac{\mathbf{J}\cdot\lhrcurs}{\lrcurs^2} d\tau' \label{eqn:cd_Lorentz}
\end{equation} 
It seems that Lorentz gauge function is not zero, which is the same situation as in classical point charge (\ref{eqn:cp_Lorentz}). In derivation of (\ref{eqn:cd_Lorentz}) the following continuity identity is used 
\begin{equation}\label{eqn:classical_continuity}
\frac{\partial{\rho}}{\partial{t}} = -\nabla \cdot \mathbf{J} \Rightarrow \dot{\rho} = \frac{\mathbf{\dot{J}} \cdot \lhrcurs}{c} 
\end{equation}
where the equation on the right is obtained if we use identity (\ref{eqn:classical_timeS}) for time differentiation and (\ref{eqn:classical_dotP}) for nabla operator. Using (\ref{eqn:cd_nablaV}), (\ref{eqn:cd_nablacrossA}), (\ref{eqn:cd_timeA}) and we can find fields
\begin{subequations}
	\begin{align}
	\mathbf{E} &= \frac{1}{4\pi\varepsilon_0}\int (\frac{\dot{\rho}\lhrcurs}{\lrcurs c} + \frac{\rho \lhrcurs}{\lrcurs^2} - \frac{\mathbf{\dot{J}}}{\lrcurs c^2}) d\tau' \label{eqn:cd_E}\\
	\mathbf{B} &= \frac{\mu_0}{4 \pi} \int (\frac{\mathbf{\dot{J}}}{\lrcurs c} + \frac{\mathbf{J}}{\lrcurs^2}) \times \lhrcurs d\tau' \label{eqn:cd_B}
	\end{align}
\end{subequations}
which are Jefimenko equations. Let's find divergence of electric and of magnetic fields using (\ref{eqn:classical_grad_power_retardedr}), (\ref{subeqn21})-(\ref{subeqn23}), (\ref{eqn:classical_dotP})-(\ref{eqn:classical_gradS})
\begin{subequations}
	\begin{align}
	\nabla \cdot \mathbf{E} &= \frac{1}{4\pi\varepsilon_0} \int (4\pi\rho\delta^3(\lbrcurs) + \frac{\mathbf{\dot{J}} \cdot \lhrcurs}{\lrcurs^2 c^2}) d\tau' \label{eqn:cd_nablaE} \\
	\nabla \cdot \mathbf{B} &= 0 \label{eqn:cd_nablaB} 
	\end{align}
\end{subequations}
In derivation of (\ref{eqn:cd_nablaE}) the continuity equation (\ref{eqn:classical_continuity}) and its time differential version are used
\begin{equation}\label{eqn:classical_continuity_diff}
\frac{\partial^2{\rho}}{\partial{t^2}} = -\frac{\partial{(\nabla \cdot \mathbf{J})}}{\partial{t}} \Rightarrow \ddot{\rho} = \frac{\mathbf{\ddot{J}} \cdot \lhrcurs}{c} 
\end{equation}
We can get also time integral version of (\ref{eqn:classical_continuity})
\begin{equation}\label{eqn:classical_continuity_int}
	\int{\frac{\partial{\rho}}{\partial{t}}} dt= -\int \nabla \cdot \mathbf{J} dt \Rightarrow \rho = \frac{\mathbf{J} \cdot \lhrcurs}{c} 
\end{equation}
which can be used to obtain the alternative version of (\ref{eqn:cd_Lorentz})
\begin{equation}\label{eqn:cd_Lorentz2}
L = -\frac{\mu_0 c}{4 \pi } \int  \frac{\rho}{\lrcurs^2} d\tau'
\end{equation} 
We can see that divergence of magnetic field for continuous charge is exactly zero. However interesting case happens for divergence of electric field, which has two components. One component is due to presence of electric charge and is similar to classical point charge result in (\ref{eqn:cp_nablaE}). The second component, however, is new and is due to time-varying electric current density. The second term implies that divergence of electric field for classical distributed charge is not always zero for volumes of space, inside of which there is no charge. In other words, in presence of time-varying distributed electric currents, divergence of electric field is not zero.  

\subsection{Relativistic continuous charge} \label{sec:relativistic_distributed_charge}

This is the last subsection. Here we keep the assumption for distributed charge, that they are functions of time $t_r$ and of space coordinates $\mathbf{r}'$. The electrodynamic potentials for relativistic distributed charge are given in the lower right corner of the Table \ref{tab:potentials}
\begin{subequations}\label{eqn:relativistic_distributed_potentials}
	\begin{align}
		V(\mathbf{r},t) = \frac{c}{4\pi \varepsilon_0} \int \frac{\rho(\mathbf{r}',t_r)}{(\lrcurs c - \lbrcurs \cdot \mathbf{v})} d\tau' \\
		\mathbf{A}(\mathbf{r},t) = \frac{\mu_0 c}{4\pi} \int \frac{\mathbf{J}(\mathbf{r}',t_r)}{(\lrcurs c - \lbrcurs \cdot \mathbf{v})} d\tau'
	\end{align}
\end{subequations}
For scalar potential, using (\ref{subeqn41}), (\ref{eqn:relativistic_dotP}), (\ref{eqn: relativistic_gradS}), (\ref{eqn:grad_term}), (\ref{eqn:grad_term2}), we have 
\begin{subequations}
	\begin{align}
		\nabla V(\mathbf{r},t) &= \frac{c}{4\pi \varepsilon_0} \int \nabla \Big(\frac{\rho}{(\lrcurs c - \lbrcurs \cdot \mathbf{v})}\Big) d\tau' \nonumber \\ 
		& = \frac{c}{4\pi \varepsilon_0} \int \Big(\frac{\rho \mathbf{v} - \dot{\rho} \lbrcurs}{(\lrcurs c - \lbrcurs \cdot \mathbf{v})^2} - \frac{\rho \lbrcurs (c^2 - \mathbf{v} \cdot \mathbf{v} + \lbrcurs \cdot \mathbf{a})}{(\lrcurs c - \lbrcurs \cdot \mathbf{v})^3}\Big) d\tau' \label{eqn:rd_nablaV}\\
		\nabla^2 V(\mathbf{r},t) &= \frac{c}{4\pi \varepsilon_0} \int \Big(\frac{1}{(\lrcurs c - \lbrcurs \cdot \mathbf{v})^3}(-\dot{\rho}(3\lrcurs c + \lbrcurs \cdot \mathbf{v})+\ddot{\rho}(\lbrcurs \cdot \lbrcurs) \nonumber \\ & - 2\rho(-\mathbf{v} \cdot \mathbf{v}+\lbrcurs \cdot \mathbf{a})) + \frac{1}{(\lrcurs c - \lbrcurs \cdot \mathbf{v})^4}(-\rho(3(\mathbf{a} \cdot \mathbf{v})-\lbrcurs \cdot \mathbf{b})(\lbrcurs \cdot \lbrcurs) \nonumber \\ 
		& + 3(c^2-\mathbf{v} \cdot \mathbf{v} + \lbrcurs \cdot \mathbf{a})(-\rho(\lrcurs c + \lbrcurs \cdot \mathbf{v})+\dot{\rho}(\lbrcurs \cdot \lbrcurs))) \nonumber \\
		& + \frac{1}{(\lrcurs c - \lbrcurs \cdot \mathbf{v})^5}(3(c^2-\mathbf{v} \cdot \mathbf{v} + \lbrcurs \cdot \mathbf{a})^2\rho(\lbrcurs \cdot \lbrcurs)) \Big) d\tau'  \label{eqn:rd_nabla2V}
	\end{align}
\end{subequations}
and using (\ref{eqn:timed_relat_retardedt}), (\ref{eqn:timed_relat_retardedr}), (\ref{eqn:time_term}), (\ref{eqn:time_term2})
\begin{subequations}
	\begin{align}
	\frac{\partial{V(\mathbf{r},t)} }{\partial{t}} &= \frac{c}{4\pi \varepsilon_0} \int \Big(\frac{\partial}{\partial{t}} \frac{\rho}{(\lrcurs c - \lbrcurs \cdot \mathbf{v})} \Big) d\tau' \nonumber \\
	& = \frac{c^2}{4\pi \varepsilon_0} \int \Big( \frac{\dot{\rho} \lrcurs - \rho c}{(\lrcurs c - \lbrcurs \cdot \mathbf{v})^2} + \frac{\rho \lrcurs (c^2-\mathbf{v} \cdot \mathbf{v} + \lbrcurs \cdot \mathbf{a})}{(\lrcurs c - \lbrcurs \cdot \mathbf{v})^3}\Big) d\tau' \label{eqn:rd_timeV}\\
	\frac{\partial^2{V(\mathbf{r},t)}}{\partial{t^2}} &= \frac{c^3}{4\pi \varepsilon_0} \int \Big( \frac{1}{(\lrcurs c - \lbrcurs \cdot \mathbf{v})^3}(-\dot{\rho}(3\lrcurs c + \lbrcurs \cdot \mathbf{v})+\ddot{\rho}(\lbrcurs \cdot \lbrcurs) \nonumber \\ & - 2\rho(-\mathbf{v} \cdot \mathbf{v}+\lbrcurs \cdot \mathbf{a}))+\frac{1}{(\lrcurs c - \lbrcurs \cdot \mathbf{v})^4}(-\rho(3(\mathbf{a} \cdot \mathbf{v})-\lbrcurs \cdot \mathbf{b})(\lbrcurs \cdot \lbrcurs) \nonumber \\ 
	& + 3(c^2-\mathbf{v} \cdot \mathbf{v} + \lbrcurs \cdot \mathbf{a})(-\rho(\lrcurs c + \lbrcurs \cdot \mathbf{v})+\dot{\rho}(\lbrcurs \cdot \lbrcurs))) \nonumber \\
	& + \frac{1}{(\lrcurs c - \lbrcurs \cdot \mathbf{v})^5}(3(c^2-\mathbf{v} \cdot \mathbf{v} + \lbrcurs \cdot \mathbf{a})^2\rho(\lbrcurs \cdot \lbrcurs)) \Big) d\tau' \label{eqn:rd_time2V}
	\end{align}
\end{subequations}
We can now check the Lorentz gauge condition for $V$ by substituting (\ref{eqn:rd_nabla2V}) and (\ref{eqn:rd_time2V}) into (\ref{eqn:Maxwell_Lorentz})
\begin{equation}
\square^2 V = 0 \label{eqn:rd_V_Lorentz}
\end{equation}
This result shows that Lorentz gauge condition for V is satisfied as for trivial case. Now we will work with vector potential, using (\ref{subeqn41})-(\ref{subeqn43}), (\ref{eqn:relativistic_dotP})-(\ref{eqn: relativistic_gradS}), (\ref{eqn:relativistic_rgradf})-(\ref{eqn:relativistic_agradb}), (\ref{eqn:grad_term}), (\ref{eqn:grad_term2}):
\begin{subequations}
	\begin{align}
	\nabla \cdot \mathbf{A} &= \frac{\mu_0c}{4 \pi} \int \nabla \cdot \Big(\frac{\mathbf{J}}{(\lrcurs c - \lbrcurs \cdot \mathbf{v})}\Big) d\tau' \nonumber \\
	& = \frac{\mu_0 c}{4 \pi} \int \Big(\frac{-\mathbf{\dot{J}} \cdot \lbrcurs + \mathbf{J} \cdot \mathbf{v}}{(\lrcurs c - \lbrcurs \cdot \mathbf{v})^2} - \frac{(c^2-\mathbf{v} \cdot \mathbf{v} + \lbrcurs \cdot \mathbf{a})(\lbrcurs \cdot \mathbf{J})}{(\lrcurs c - \lbrcurs \cdot \mathbf{v})^3}\Big) d\tau' \label{eqn:rd_nabladotA} \\
	\nabla(\nabla \cdot \mathbf{A}) &= \frac{\mu_0c}{4 \pi} \int \Big( \frac{1}{(\lrcurs c - \lbrcurs \cdot \mathbf{v})^2}(-\mathbf{\dot{J}}) + \frac{1}{(\lrcurs c - \lbrcurs \cdot \mathbf{v})^3}(\lbrcurs(\lbrcurs \cdot \mathbf{\ddot{J}})-2\lbrcurs(\mathbf{\dot{J}} \cdot \mathbf{v}) \nonumber \\
	& -2 \mathbf{v}(\lbrcurs \cdot \mathbf{\dot{J}})-\mathbf{J}(c^2-\mathbf{v} \cdot \mathbf{v} + \lbrcurs \cdot \mathbf{a}) -\lbrcurs(\mathbf{J} \cdot \mathbf{a}) + 2\mathbf{v}(\mathbf{J} \cdot \mathbf{v}) - \mathbf{a}(\lbrcurs \cdot \mathbf{J})) \nonumber \\
	& + \frac{1}{(\lrcurs c - \lbrcurs \cdot \mathbf{v})^4}(-\lbrcurs(\lbrcurs \cdot \mathbf{J})(3(\mathbf{a} \cdot \mathbf{v}) - \lbrcurs \cdot \mathbf{b}) +3(c^2-\mathbf{v} \cdot \mathbf{v} + \lbrcurs \cdot \mathbf{a})(\lbrcurs(\lbrcurs \cdot \mathbf{\dot{J}}) \nonumber \\
	& -\lbrcurs(\mathbf{J} \cdot \mathbf{v})-\mathbf{v}(\lbrcurs \cdot \mathbf{J}))) + \frac{1}{(\lrcurs c - \lbrcurs \cdot \mathbf{v})^5}(3\lbrcurs(\lbrcurs \cdot \mathbf{J})(c^2-\mathbf{v} \cdot \mathbf{v} + \lbrcurs \cdot \mathbf{a})^2) \Big) d\tau' \label{eqn:rd_nabladot2A} \\
	\nabla \times \mathbf{A} &= \frac{\mu_0c}{4 \pi} \int \nabla \times \Big(\frac{\mathbf{J}}{(\lrcurs c - \lbrcurs \cdot \mathbf{v})}\Big) d\tau' \nonumber \\ 
	& = \frac{\mu_0c}{4 \pi} \int \Big(\frac{\mathbf{\dot{J}} \times \lbrcurs - \mathbf{J} \times \mathbf{v}}{(\lrcurs c - \lbrcurs \cdot \mathbf{v})^2} + \frac{(\mathbf{J} \times \lbrcurs)(c^2-\mathbf{v} \cdot \mathbf{v} + \lbrcurs \cdot \mathbf{a})}{(\lrcurs c - \lbrcurs \cdot \mathbf{v})^3}\Big) d\tau' \label{eqn:rd_nablacrossA} \\
	\nabla \times (\nabla \times \mathbf{A}) &= \frac{\mu_0c}{4 \pi} \int \Big( \frac{1}{(\lrcurs c - \lbrcurs \cdot \mathbf{v})^3}(-\mathbf{\ddot{J}}(\lbrcurs \cdot \lbrcurs)+\lbrcurs(\lbrcurs \cdot \mathbf{\ddot{J}})-2\mathbf{v}(\lbrcurs \cdot \mathbf{\dot{J}})-2\lbrcurs(\mathbf{\dot{J}} \cdot \mathbf{v}) \nonumber \\
	& + 2\mathbf{\dot{J}}(\lrcurs c + \lbrcurs \cdot \mathbf{v})-2\mathbf{J}(\mathbf{v} \cdot \mathbf{v})+2\mathbf{v}(\mathbf{J} \cdot \mathbf{v})-\mathbf{a}(\mathbf{J} \cdot \lbrcurs)+2\mathbf{J}(\lbrcurs \cdot \mathbf{a})-\lbrcurs(\mathbf{J} \cdot \mathbf{a})) \nonumber \\
	& + \frac{1}{(\lrcurs c - \lbrcurs \cdot \mathbf{v})^4}((c^2-\mathbf{v} \cdot \mathbf{v} + \lbrcurs \cdot \mathbf{a})(3(\lbrcurs(\mathbf{\dot{J}} \cdot \lbrcurs)-\mathbf{\dot{J}}(\lbrcurs \cdot \lbrcurs))+4\mathbf{J}(\lbrcurs \cdot \mathbf{v}) \nonumber \\
	& -3\mathbf{v}(\mathbf{J} \cdot \lbrcurs)+2\lrcurs c \mathbf{J} - 3\lbrcurs(\mathbf{J} \cdot \mathbf{v}))+(3(\mathbf{a} \cdot \mathbf{v})-\lbrcurs \cdot \mathbf{b})(\mathbf{J}(\lbrcurs \cdot \lbrcurs)-\lbrcurs(\mathbf{J} \cdot \lbrcurs))) \nonumber \\ 
	&  + \frac{1}{(\lrcurs c - \lbrcurs \cdot \mathbf{v})^5}(3(-\mathbf{J}(\lbrcurs \cdot \lbrcurs) + \lbrcurs (\mathbf{J} \cdot \lbrcurs))(c^2-\mathbf{v} \cdot \mathbf{v} + \lbrcurs \cdot \mathbf{a})^2) \Big) d\tau' \label{eqn:rd_nablacross2A}
	\end{align}
\end{subequations}
Using (\ref{eqn:timed_relat_retardedt}), (\ref{eqn:timed_relat_retardedr}), (\ref{subeqn31}), (\ref{eqn:relativistic_time_power_retardedr}), (\ref{subeqn44}), (\ref{eqn:relativistic_timeP}), (\ref{eqn:relativistic_timeS}), (\ref{eqn:time_term}), (\ref{eqn:time_term2}) we find time derivatives 
\begin{subequations}
	\begin{align}
	\frac{\partial{\mathbf{A}}}{\partial{t}} &= \frac{\mu_0c}{4\pi} \int \Big(\frac{\partial}{\partial{t}} \frac{\mathbf{J}}{(\lrcurs c - \lbrcurs \cdot \mathbf{v})} \Big) d\tau' \nonumber \\
	& = \frac{\mu_0c^2}{4\pi} \int \Big(\frac{\mathbf{\dot{J}} \lrcurs - \mathbf{J} c}{(\lrcurs c - \lbrcurs \cdot \mathbf{v})^2} + \frac{\mathbf{J} \lrcurs (c^2-\mathbf{v} \cdot \mathbf{v} + \lbrcurs \cdot \mathbf{a})}{(\lrcurs c - \lbrcurs \cdot \mathbf{v})^3}\Big) d\tau' \label{eqn:rd_timeA}\\
	\frac{\partial^2{\mathbf{A}}}{\partial{t^2}} &= \frac{\mu_0c^3}{4\pi} \int \Big( \frac{1}{(\lrcurs c - \lbrcurs \cdot \mathbf{v})^3}(-\mathbf{\dot{J}}(\lbrcurs \cdot \mathbf{v}) + \mathbf{\ddot{J}}(\lbrcurs \cdot \lbrcurs) - 3 \lrcurs c \mathbf{\dot{J}} + 2c^2 \mathbf{J} \nonumber \\
	& - 2\mathbf{J}(c^2-\mathbf{v} \cdot \mathbf{v} + \lbrcurs \cdot \mathbf{a}))+\frac{1}{(\lrcurs c - \lbrcurs \cdot \mathbf{v})^4}(-\mathbf{J}(3(\mathbf{a} \cdot \mathbf{v})-\lbrcurs \cdot \mathbf{b})(\lbrcurs \cdot \lbrcurs) \nonumber \\ 
	& + 3(c^2-\mathbf{v} \cdot \mathbf{v} + \lbrcurs \cdot \mathbf{a})(\mathbf{\dot{J}}(\lbrcurs \cdot \lbrcurs)-\lrcurs c \mathbf{J} - \mathbf{J}(\lbrcurs \cdot \mathbf{v}))) \nonumber \\
	& + \frac{1}{(\lrcurs c - \lbrcurs \cdot \mathbf{v})^5}(3(c^2-\mathbf{v} \cdot \mathbf{v} + \lbrcurs \cdot \mathbf{a})^2\mathbf{J}(\lbrcurs \cdot \lbrcurs)) \Big) d\tau' \label{eqn:rd_time2A}
	\end{align}
\end{subequations}
Let's put (\ref{eqn:rd_nabladot2A}), (\ref{eqn:rd_nablacross2A}) and (\ref{eqn:rd_time2A}) into (\ref{eqn:Maxwell_Lorentz})
 \begin{equation}
 \square^2 \mathbf{A} = 0 \label{eqn:rd_A_Lorentz}
 \end{equation}
 For vector potential Lorentz gauge condition is trivially satisfied. Lorentz gauge function for relativistic distributed charge is 
 \begin{equation}
 L = \nabla \cdot \mathbf{A} + \mu_0 \varepsilon_{0} \frac{\partial{V}}{\partial{t}} = \frac{\mu_0c}{4 \pi } \int  \Big( \frac{\mathbf{J}\cdot\mathbf{v}-\rho c^2}{(\lrcurs c - \lbrcurs \cdot \mathbf{v})^2} + \frac{(\lrcurs c \rho - \lbrcurs \cdot \mathbf{J})(c^2-\mathbf{v} \cdot \mathbf{v} + \lbrcurs \cdot \mathbf{a})}{(\lrcurs c - \lbrcurs \cdot \mathbf{v})^3} \Big) d\tau' \label{eqn:rd_Lorentz}
 \end{equation} 
 We can see that Lorentz gauge function is not zero and Lorentz gauge condition is not satisfied for relativistic continuous charge. As in the case for classical continuous charge, in derivation of (\ref{eqn:rd_Lorentz}) we used continuity equation 
  \begin{equation}\label{eqn:relativistic_continuity}
  \frac{\partial{\rho}}{\partial{t}} = -\nabla \cdot \mathbf{J} \Rightarrow \lrcurs c \dot{\rho} = \mathbf{\dot{J}} \cdot \lbrcurs 
  \end{equation}
The equation on right side of (\ref{eqn:relativistic_continuity}) is obtained if (\ref{eqn:relativistic_timeS}) is used for time differentiation and (\ref{eqn:relativistic_dotP}) is used for divergence operator. 
 Let's now find fields using (\ref{eqn:rd_nablaV}), (\ref{eqn:rd_nablacrossA}) and (\ref{eqn:rd_timeA})
 \begin{subequations}
 	\begin{align}
 	\mathbf{E} &= \frac{1}{4\pi\varepsilon_0}\int \Big(\frac{1}{(\lrcurs c - \lbrcurs \cdot \mathbf{v})^2}(\dot{\rho}c\lbrcurs - \rho c\mathbf{v} - \lrcurs \mathbf{\dot{J}} + c\mathbf{J}) \nonumber \\
 	& + \frac{1}{(\lrcurs c - \lbrcurs \cdot \mathbf{v})^3}((\rho c \lbrcurs - \lrcurs \mathbf{J})(c^2-\mathbf{v} \cdot \mathbf{v} + \lbrcurs \cdot \mathbf{a})) \Big) d\tau' \label{eqn:rd_E}\\
 	\mathbf{B} &= \frac{\mu_0 c}{4\pi}\int \Big(\frac{1}{(\lrcurs c - \lbrcurs \cdot \mathbf{v})^2}(\mathbf{\dot{J}} \times \lbrcurs - \mathbf{J} \times \mathbf{v}) \nonumber \\
 	& + \frac{1}{(\lrcurs c - \lbrcurs \cdot \mathbf{v})^3}((\mathbf{J} \times \lbrcurs)(c^2-\mathbf{v} \cdot \mathbf{v} + \lbrcurs \cdot \mathbf{a})) \Big) d\tau' \label{eqn:rd_B}
 	\end{align}
 \end{subequations}
 These fields are relativistic versions of Jefimenko equations. Finally, we are ready to find divergence of electric and of magnetic fields
 \begin{subequations}
 	\begin{align}
 	\nabla \cdot \mathbf{E} &= \frac{1}{4\pi\varepsilon_0}\int \Big(\frac{1}{(\lrcurs c - \lbrcurs \cdot \mathbf{v})^3}(\dot{\rho}\lrcurs c^2 + 2 \rho c(-\mathbf{v} \cdot \mathbf{v} + \lbrcurs \cdot \mathbf{a}) - \lrcurs \mathbf{\dot{J}} \cdot \mathbf{v} \nonumber \\
 	& + 2c \mathbf{J}\cdot\mathbf{v} - \lrcurs \mathbf{a} \cdot \mathbf{J}) + \frac{1}{(\lrcurs c - \lbrcurs \cdot \mathbf{v})^4}((3\mathbf{a} \cdot \mathbf{v} - \lbrcurs \cdot \mathbf{b})(\rho c \lrcurs^2 - \lrcurs (\lbrcurs \cdot \mathbf{J})) \nonumber \\
 	& + (3\rho c(\lrcurs c + \lbrcurs \cdot \mathbf{v})-3c \mathbf{J} \cdot \lbrcurs - 3\lrcurs(\mathbf{J} \cdot \mathbf{v}))(c^2-\mathbf{v} \cdot \mathbf{v} + \lbrcurs \cdot \mathbf{a})) \nonumber \\
 	& + \frac{1}{(\lrcurs c - \lbrcurs \cdot \mathbf{v})^5}(-3c\rho \lrcurs^2+3\lrcurs \mathbf{J} \cdot \lbrcurs)(c^2-\mathbf{v} \cdot \mathbf{v} + \lbrcurs \cdot \mathbf{a})^2 \Big) d\tau' \label{eqn:rd_nablaE} \\
 	\nabla \cdot \mathbf{B} &= 0 \label{eqn:rd_nablaB} 
 	\end{align}
 \end{subequations}
 In derivation of (\ref{eqn:rd_nablaE}) the continuity equation (\ref{eqn:relativistic_continuity}) and its time differential version are used 
 \begin{equation}\label{eqn:relativistic_continuity_diff}
\frac{\partial^2{\rho}}{\partial{t^2}} = -\frac{\partial{(\nabla \cdot \mathbf{J})}}{\partial{t}} \Rightarrow -c(\lbrcurs \cdot \mathbf{v}) \dot{\rho} + \lrcurs^2 c \ddot{\rho} = -\lrcurs (\mathbf{v} \cdot \mathbf{\dot{J}}) + \lrcurs (\lbrcurs \cdot \mathbf{\ddot{J}})
 \end{equation}
 In order to obtain the time integral version of (\ref{eqn:relativistic_continuity}), it needs to be integrated with respect to time. However at this point in time we could not find analytical expression for integral of (\ref{eqn:relativistic_continuity}), because $\lrcurs$ and $\lbrcurs$ are functions of time also, which we cannot neglect. 
 
We can see that divergence of electric field for continuous charge in relativistic case is not zero. While divergence of magnetic field is exactly zero.
 
\section{Discussion}

Divergence of magnetic field for all four cases considered here is exactly zero, see (\ref{eqn:cp_nablaB}), (\ref{eqn:rp_nablaB}), (\ref{eqn:cd_nablaB}) and (\ref{eqn:rd_nablaB}). This result is expected, because by definition vector potential was introduced using vector calculus identity that divergence of curl of any continuous vector field is exactly zero (\ref{eqn:Gauss2}). However interesting case takes place for divergence of electric field. For classical point charge it is proportional to net sum of charges present inside of volume under consideration (\ref{eqn:cp_nablaE}). Therefore for classical point charge Gauss's law is still valid. For relativistic point charge divergence of electric field is exactly zero (\ref{eqn:rp_nablaE}), no matter how it moves. We note that result (\ref{eqn:rp_nablaE}) fully agrees with recent discussion of electric field of a point charge in truncated hyperbolic motion, where it is shown that Gauss's law is still valid and that long-standing apparent violation of the Gauss's law is not correct \cite{Griffiths2014, Franklin2015}. For classical continuous charge it is not zero (\ref{eqn:cd_nablaE}). More interestingly, it has two terms. One term is proportional to net sum of charges present inside of volume under consideration, which is the same as for classical point charge. However the second terms has time derivative of current density, or if we use (\ref{eqn:classical_continuity}), then (\ref{eqn:cd_nablaE}) can be rewritten as 
\begin{equation*}
\nabla \cdot \mathbf{E} = \frac{1}{4\pi\varepsilon_0} \int (4\pi\rho\delta^3(\lbrcurs) + \frac{\dot{\rho}}{\lrcurs^2 c}) d\tau'
\end{equation*}
which means that divergence of electric field is not zero if continuous charge density is time-varying. For example, during ionization (charging) or recombination (discharging) process, when net electrical charge of a system varies with time, divergence of electric field will also vary with time. In other words, this result shows that divergence of time-varying electric field, even outside of charges, which generated that field, might be non-zero. The fact that the second terms has $c$ or $c^2$ in the denominator might suggest that this term has much smaller effect than the first term. In practice this might explain why this effect was not detected experimentally up to this point, especially for slowly varying electric charges or currents. However because we have time derivative in the numerator, very-fast changing electric charges or currents might generate sufficiently large effects of variations in divergence of electric field, which might be detected experimentally. Also from (\ref{eqn:cd_nablaE}) we can see that the second term has square dependence on distance in the denominator. Thus we can anticipate that divergence of electric field is a scalar field, which falls of with distance in the same manner as electric field does. In other words, divergence of electric field might be far-type field.

From (\ref{eqn:cd_nablaE}) it follows that if $\mathbf{\ddot{J}}$ and $\lhrcurs$ are perpendicular to each other, then the second terms is zero, while if they are aligned, then the second terms has maximum value. This suggests that the effect of this term is direction-dependent or anisotropic. We should give one word of caution. Time-varying electric currents in (\ref{eqn:cd_nablaE}) can be generated either by a constant charge, which moves with acceleration, or by a time-varying charge, which might be stationary or it might be moving arbitrarily. The former case corresponds to (\ref{eqn:rp_nablaE}), and so for this case divergence of electric field is zero. The latter case corresponds to (\ref{eqn:cd_nablaE}) or to (\ref{eqn:rd_nablaE}), and so divergence of electric field is not zero. In other words, time-varying electric currents in (\ref{eqn:cd_nablaE}) and in (\ref{eqn:rd_nablaE}) are generated by time-varying charges, according to (\ref{eqn:classical_continuity}) and (\ref{eqn:relativistic_continuity}).

Finally, for relativistic distributed charge the divergence of electric field is not zero (\ref{eqn:rd_nablaE}). Divergence has some complex terms. However the main result is the same, which is that divergence of electric field, generated by time-varying sources (charges or currents), is not zero, even for volumes of space, which do not contain these sources. Therefore we can conclude that for time-varying electric charges and currents Gauss's law needs to be modified. We believe that these results were escaping attention of scientists before because in most textbooks and papers only motion of a point charge is considered. As we have seen, charge value of a point charge does not change and is constant, and so divergence of electric field satisfies original Gauss' law, i.e. it is either constant or zero. However for continuous charge its charge value changes as a function of time and space, and so original Gauss' law does not hold and divergence of electric field itself becomes function of space and time. This result becomes even more clear if we consider the following ``Gedanken'' experiment. Imagine closed volume, isolated electrically from outside universe, with net zero charge inside at initial time. Therefore divergence of electric field is zero at initial time. Imagine that at some later time due to ionization and/or charging net nonzero charge is created inside. As this charge appears, its electric field lines start to spread in space of the isolated volume. The speed, with which electric field lines spread in space is finite, and so divergence of electric field will change from initial zero value to some nonzero value at later time. 

We note that results for relativistic cases can be reduced to their corresponding classical cases by simply setting $\mathbf{v}$ and $\mathbf{a}$ equal to $\mathbf{0}$. It is also interesting that no boost $\mathbf{b}$ is present in the equations for $\mathbf{E}$ and for $\mathbf{B}$ and for their divergences.  

\section{Conclusion}

In this paper we considered divergence of electric and of magnetic fields for four cases: classical point charge, classical continuous charge, relativistic point and relativistic continuous charges. Results for classical and relativistic point charges are the same as in literature, i.e. Gauss's law is valid. However results for time-varying classical and relativistic distributed charges indicate that divergence of electric field is not zero even for volumes of space where no charges are present. For these cases original Gauss's law might require modification. Divergence of electric field seems to be far-field type scalar anisotropic field, which is generated by time-varying electric charges or currents. Results indicate that for these effects to be sufficiently large to be experimentally observable the time variation of electric charges and/or of currents should be very fast. Divergence of magnetic field is zero for all cases.

\bibliographystyle{IEEEtran}
\bibliography{bibliography4}

\begin{thebibliography}{1}
\providecommand{\url}[1]{#1}
\csname url@samestyle\endcsname
\providecommand{\newblock}{\relax}
\providecommand{\bibinfo}[2]{#2}
\providecommand{\BIBentrySTDinterwordspacing}{\spaceskip=0pt\relax}
\providecommand{\BIBentryALTinterwordstretchfactor}{4}
\providecommand{\BIBentryALTinterwordspacing}{\spaceskip=\fontdimen2\font plus
\BIBentryALTinterwordstretchfactor\fontdimen3\font minus
  \fontdimen4\font\relax}
\providecommand{\BIBforeignlanguage}[2]{{%
\expandafter\ifx\csname l@#1\endcsname\relax
\typeout{** WARNING: IEEEtran.bst: No hyphenation pattern has been}%
\typeout{** loaded for the language `#1'. Using the pattern for}%
\typeout{** the default language instead.}%
\else
\language=\csname l@#1\endcsname
\fi
#2}}
\providecommand{\BIBdecl}{\relax}
\BIBdecl

\bibitem{DavidGriffiths}
D.~J. Griffiths, \emph{Introduction to Electrodynamics}.\hskip 1em plus 0.5em
  minus 0.4em\relax USA: Prentice Hall, 1999.

\bibitem{WolfgangPanofsky}
W.~K.~H. Panofsky and M.~Phillips, \emph{Classical Electricity and Magnetism},
  2nd~ed.\hskip 1em plus 0.5em minus 0.4em\relax USA: Addison-Wesley Publishing
  Company, Inc., 1962.

\bibitem{JerroldFranklin}
J.~Franklin, \emph{Classical Electromagnetism}.\hskip 1em plus 0.5em minus
  0.4em\relax USA: Pearson, 2005.

\bibitem{WalterGreiner}
W.~Greiner, \emph{Classical Electrodynamics}.\hskip 1em plus 0.5em minus
  0.4em\relax Springer, 1998.

\bibitem{Griffiths2014}
J.~Franklin and D.~J. Griffiths, ``The fields of a charged particle in
  hyperbolic motion,'' \emph{American Journal of Physics}, vol.~82, no.~8, pp.
  755--763, 2014.

\bibitem{Franklin2015}
J.~Franklin, ``Electric field of a point charge in truncated hyperbolic
  motion,'' \emph{European Journal of Physics}, vol.~36, pp. 1--8, 2015.

\end{thebibliography}

\end{document}